\documentclass[sigconf]{acmart}
\usepackage{balance}
\AtBeginDocument{%
  \providecommand\BibTeX{{%
    \normalfont B\kern-0.5em{\scshape i\kern-0.25em b}\kern-0.8em\TeX}}}

\setcopyright{acmcopyright}
\copyrightyear{2020}
\acmYear{2020}
\acmDOI{10.1145/3340531.3412880}

\acmConference[CIKM '20]{Proceedings of the 29th ACM International Conference on Information and Knowledge Management}{October 19--23, 2020}{Virtual Event, Ireland}
\acmBooktitle{Proceedings of the 29th ACM International Conference on Information and Knowledge Management (CIKM '20), October 19--23, 2020, Virtual Event, Ireland}
\acmPrice{15.00}
\acmISBN{978-1-4503-6859-9/20/10}

\usepackage{amsmath,amssymb,amsfonts}
\usepackage{algorithmic}
\usepackage{graphicx}
\usepackage{textcomp}
\usepackage{enumitem}
\usepackage{subfigure}
\usepackage{adjustbox}
\usepackage{array, boldline, makecell, booktabs}
\usepackage{epsfig}
\usepackage[mathscr]{euscript}
\usepackage{longtable}
\usepackage{lscape}
\usepackage{float}
\usepackage{multirow}
\usepackage[figuresright]{rotating}
\usepackage[para,online,flushleft]{threeparttable}
\usepackage{tabularx}
\usepackage{url}
\usepackage[boxed,ruled,commentsnumbered]{algorithm2e}
\usepackage{multirow}
\usepackage{xcolor}
\usepackage{diagbox}
\usepackage{url}

\begin{document}
\fancyhead{}

\title{$\mathsf{ReCOVery}$: A Multimodal Repository for COVID-19 News Credibility Research}

\author{Xinyi Zhou}
\email{zhouxinyi@data.syr.edu}
\orcid{0002-2388-254X}
\affiliation{%
  \institution{Data Lab, EECS Department\\Syracuse University}
}
\author{Apurva Mulay}
\email{asmulay@syr.edu}
\affiliation{%
  \institution{Data Lab, EECS Department\\Syracuse University}
}
\author{Emilio Ferrara}
\email{emiliofe@usc.edu}
\affiliation{%
  \institution{Information Sciences Institute, \\University of Southern California}
}
\author{Reza Zafarani}
\email{reza@data.syr.edu}
\affiliation{%
  \institution{Data Lab, EECS Department\\Syracuse University}
}


\begin{abstract}
First identified in Wuhan, China, in December 2019, the outbreak of COVID-19 has been declared as a global emergency in January, and a pandemic in March 2020 by the World Health Organization (WHO). Along with this pandemic, we are also experiencing an ``infodemic'' of information with low credibility such as fake news and conspiracies. In this work, we present $\mathsf{ReCOVery}$, a repository designed and constructed to facilitate research on combating such information regarding COVID-19. We first broadly search and investigate $\sim$2,000 news publishers, from which 60 are identified with extreme [high or low] levels of credibility. By inheriting the credibility of the media on which they were published, a total of 2,029 news articles on coronavirus, published from January to May 2020, are collected in the repository, along with 140,820 tweets that reveal how these news articles have spread on the Twitter social network. The repository provides multimodal information of news articles on coronavirus, including textual, visual, temporal, and network information. The way that news credibility is obtained allows a trade-off between dataset scalability and label accuracy. Extensive experiments are conducted to present data statistics and distributions, as well as to provide baseline performances for predicting news credibility so that future methods can be compared. Our repository is available at \url{http://coronavirus-fakenews.com}.
\end{abstract}



\maketitle

\section{Introduction}

As of June $4^{\text{th}}$, the COVID-19 pandemic has resulted in over 6.4 million confirmed cases and over 380,000 deaths globally.\footnote{\url{https://www.who.int/docs/default-source/coronaviruse/situation-reports/20200604-covid-19-sitrep-136.pdf}\label{footnote:who_report}} Governments have enforced border shutdowns, travel restrictions, and quarantines to ``flatten the curve''.
The COVID-19 outbreak has had a detrimental impact on not only the healthcare sector but also every aspect of human life such as education and economic sectors~\cite{nicola2020socio}. For example, over 100 countries have imposed nationwide (even complete) closures of education facilities, which has lead to over 900 million learners being affected.\footnote{\url{https://en.unesco.org/covid19/educationresponse}} Statistics indicate that 3.3 million Americans applied for unemployment benefits in the week ending on March $21^{\text{th}}$ and the number doubled in the following week, before which the highest number of unemployment applications ever received in one week was 695,000 in 1982.\footnote{\url{https://www.npr.org/2020/03/26/821580191/unemployment-claims-expected-to-shatter-records}}

Along with the COVID-19 pandemic, we are also experiencing an ``infodemic'' of information with low credibility regarding COVID-19.\footnote{\url{https://www.un.org/en/un-coronavirus-communications-team/un-tackling-\%E2\%80\%98infodemic\%E2\%80\%99-misinformation-and-cybercrime-covid-19}\label{footnote:un_news}} Hundreds of news websites have contributed to publishing false coronavirus information.\footnote{\url{https://www.newsguardtech.com/coronavirus-misinformation-tracking-center/}\label{footnote:newsguard}} Individuals who believe false news articles (e.g., claiming that eating boiled garlic or drinking chlorine dioxide, an industrial bleach, can cure or prevent coronavirus), might take ineffective or extremely dangerous actions to protect themselves from the virus.\footnote{\url{https://www.factcheck.org/2020/02/fake-coronavirus-cures-part-1-mms-is-industrial-bleach/}\label{footnote:fake_news}}


Given this background, research is motivated to combat this infodemic. Hence, we design and construct a multimodal repository, $\mathsf{ReCOVery}$, to facilitate reliability assessment of news on COVID-19. We first broadly search and investigate $\sim$2,000 news publishers, from which 60 with various political polarizations and from different countries are identified with extreme [high or low] credibility. As past literature has indicated, there is a close relationship between the credibility of news articles and their publication sources~\cite{reza2019tutorial}. In total, 2,029 news articles on coronavirus are finally collected in the repository along with 140,820 tweets that reveal how these news articles are spread on the social network. The main contributions of this work are summarized as follows:
\begin{enumerate}
    \item We construct a repository to support the research that investigates (i)~how news with low credibility is created and spread in the COVID-19 pandemic and (ii)~ways to predict such ``fake'' news. The manner in which the ground truth of news credibility is obtained allows a scalable repository, as annotators need not label each news article that is time-consuming and instead they can directly label the news site;
    \item $\mathsf{ReCOVery}$ provides multimodal information on COVID-19 news articles. For each news article, we collect its news content and social information revealing how it spreads on social media, which covers textual, visual, temporal, and network information; and
    \item We conduct extensive experiments using $\mathsf{ReCOVery}$ data, which includes data analyses (data statistics and distributions) and baseline performances for predicting news credibility. These baselines allow future methods to be compared to. Baselines are obtained using either single-modal or multi-modal information of news articles and utilize either traditional statistical learning or deep learning. 
\end{enumerate}

The rest of this paper is organized as follows. We first review the related datasets in Section~\ref{sec:related_work}. Then, we detail how the data is collected in Section~\ref{sec:data_collection}. The statistics and distributions of the data are presented and analyzed in Section~\ref{sec:data_analysis}. Experiments that use the data to predict news credibility are designed and conducted in Section~\ref{sec:experiments}, whose results can be used as benchmarks. We conclude in Section~\ref{sec:conclusion}.

\begin{figure*}
    \includegraphics[width=0.85\textwidth]{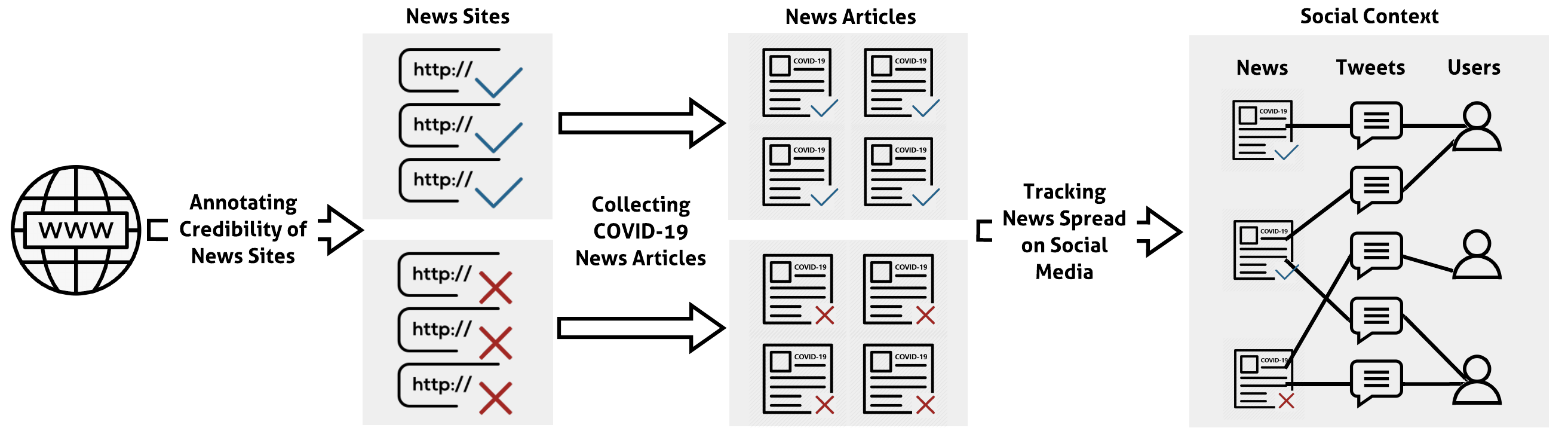}
    \caption{Data Collection Process for $\mathsf{ReCOVery}$}
    \label{fig:data_collection}
\end{figure*}

\section{Related Work}
\label{sec:related_work}

Related datasets can be generally grouped as (I) COVID-19 datasets and (II) ``fake'' news and rumor datasets.

\paragraph{COVID-19 Datasets}  
As a global emergency, the outbreak of COVID-19 has been labelled as a black swan event and likened to the economic scene of World War II~\cite{nicola2020socio}. With this background, a group of datasets have emerged, whose contributions range from real-time tracking of COVID-19 to help epidemiological forecasting (e.g., \cite{dong2020interactive} and \cite{xu2020epidemiological}) and collecting scholarly COVID-19 articles for literature-based discoveries (e.g., CORD-19\footnote{\url{https://www.semanticscholar.org/cord19}\label{footnote:cord19}}), to tracking the spreading of COVID-19 information on Twitter (e.g., \cite{chen2020tracking}).

Specifically, researchers at Johns Hopkins University develop a Web-based dashboard\footnote{\url{https://coronavirus.jhu.edu/map.html}} to visualize and track reported cases of COVID-19 in real-time. The dashboard is released on January $22^{\text{nd}}$, presenting the location and number of confirmed COVID-19 cases, deaths, and recoveries for all affected countries~\cite{dong2020interactive}. Another dataset shared publicly on March $24^{\text{th}}$ is constructed to aid the analysis and tracking of the COVID-19 epidemic, which provides real-time individual-level data (e.g., symptoms; date of onset, admission, and confirmation; and travel history) from national, provincial, and municipal health reports~\cite{xu2020epidemiological}. 
The Allen Institute for AI has contributed a free and dynamic database of more than 128,000 scholarly articles about COVID-19, named CORD-19, to the global research community.\textsuperscript{\ref{footnote:cord19}} The intention is to mobilize researchers to apply recent advances in Natural Language Processing (NLP) to generate new insights to support the fight against COVID-19. Furthemore, Chen et al.~\cite{chen2020tracking} release the first large-scale COVID-19 twitter dataset. The dataset, updated regularly, collects COVID-19 tweets that are posted from January $21^{\text{st}}$ and across languages.  

Though these datasets have been broadly investigated and have contributed to the research on coronavirus pandemic, they do not provide the ground truth on the credibility of information on coronavirus to help fight the coronavirus
infodemic.

\paragraph{``Fake'' News and Rumor Datasets}
Existing ``fake'' news and rumor datasets are collected with various focuses. These datasets may 
(i)~only contain news content that can be full articles (e.g., NELA-GT-2018~\cite{norregaard2019nela} or short claims (e.g., FEVER~\cite{thorne2018fever}); 
(ii)~only contain social media information (e.g., CREDBANK\cite{mitra2015credbank}), where news refers to user posts; or 
(iii)~contain both content and social media information (e.g., LIAR~\cite{wang2017liar}, FakeNewsNet~\cite{shu2018fakenewsnet}, and FakeHealth~\cite{dai2020ginger}).

Specifically, NELA-GT-2018~\cite{norregaard2019nela} is a large-scale dataset of around 713,000 news articles from February to November 2018. News articles are collected from 194 news medium with multiple labels directly obtained from NewsGuard, Pew Research Center, Wikipedia, OpenSources, MBFC, AllSides, BuzzFeed News, and PolitiFact. These labels refer to news credibility, transparency, political polarizations, and authenticity. 
FEVER dataset~\cite{thorne2018fever} consists of $\sim$185,000 claims and is constructed following two steps: claim generation and annotation. First, the authors extract sentences from Wikipedia, and then the annotators manually generate a set of claims based on the extracted sentences. Then, the annotators label each claim as ``supported'', ``refuted'', or ``not enough information'' by comparing it with the original sentence from which it is developed. 
On the other hand, some datasets focus on user posts on social media, for example, CREDBANK~\cite{mitra2015credbank} includes more than 60 million tweets grouped into 1049 real-world events, each of which is annotated by 30 human annotators, while some contain both news content and social media information. By collecting both claims and fact-check results (labels, i.e., ``true'', ``mostly true'', ``half-true'', ``mostly false'', and ``pants on fire'') directly from PolitiFact, Wang establishes the LIAR dataset~\cite{wang2017liar} containing around 12,800 verified statements made in public speeches and social medium. The aforementioned datasets only contain textual information valuable for NLP research with limited information on how ``fake'' news and rumors spread on social networks, which motivate the construction of FakeNewsNet and FakeHealth dataset~\cite{shu2018fakenewsnet,dai2020ginger}. The FakeNewsNet dataset collects fact-checked (real or fake) full news articles from PolitiFact (\#=1,056) and GossipCop (\#=22,140), respectively and tracks news spreading on Twitter. The FakeHealth dataset collects verified (real or fake) news reviews from \url{HealthNewsReview.org} with detailed explanations and social engagements regarding news spreading on Twitter that includes a user-user social network. Note that FakeHealth concentrates on healthcare data, so does CoAID, a recently released dataset for COVID-19 misinformation research~\cite{cui2020coaid}. 

In general, compared to datasets such as NELA-GT-2018, FEVER, and LIAR, our repository provides multimodal information and social engagements of news articles. Compared to CREDBANK and FakeNewsNet, ReCOVery aims to fight the coronavirus infodemic and presents a novel approach to collecting and annotating data, which allows the trade-off between data scalability and label accuracy. Compared to FakeHealth and CoAID, news articles in ReCOVery are from a mix of domains that include healthcare.

\section{Data Collection}
\label{sec:data_collection}

The overall process that we collect the data, including news content and social media information, is presented in Figure~\ref{fig:data_collection}. To facilitate scalability, news credibility is assessed based on the credibility of the media (site) that publishes the news article. Based on the process outlined in Figure~\ref{fig:data_collection}, we will further detail how the data is collected, answering the following three questions: (1)~how to identify reliable (or unreliable) news sites mainly releasing real news (or fake news)? (which we address in Section~\ref{subsec:news_sites}); having determined such news sites, (2) how do we crawl COVID-19 news articles from these sites and what news components are valuable for collection? (Section~\ref{subsec:news_content}); and given COVID-19 news articles, (3) how can we track their spread on social networks? (Section~\ref{subsec:social_media})

\subsection{Filtering News Sites}
\label{subsec:news_sites}

To determine a list of reliable and unreliable news sites, we primarily rely on two resources: \textit{NewsGuard} and \textit{Media Bias/Fact Check}.

\paragraph{NewsGuard}\footnote{\url{https://www.newsguardtech.com/}}
NewsGuard is developed to review and rate news websites. Its reliability rating team is formed by trained journalists and experienced editors, whose credentials and backgrounds are all transparent and available on the site. The performance (credibility) of each news website is assessed based on the following nine journalistic criteria:
\vspace{-1mm}
\begin{enumerate}
    \item Does not repeatedly publish false content,\hfill (22 points)
    \item Gathers and presents information responsibly,\hfill (18 points)
    \item Regularly corrects or clarifies errors,\hfill (12.5 points)
    \item Handles the difference between news and opinion responsibly,\hfill(12.5~points)
    \item Avoids deceptive headlines,\hfill (10 points)
    \item Website discloses ownership and financing,\hfill (7.5 points)
    \item Clearly labels advertising,\hfill (7.5 points)
    \item Reveals who’s in charge, including possible conflicts of interest, and\hfill (5~points)
    \item The site provides the names of content creators, along with either contact or biographical information, \hfill (5 points)
\end{enumerate}
\vspace{-1mm}
\noindent where the overall score of a site is between 0 to 100; 0 indicates the lowest credibility, and 100 indicates the highest credibility. A news website with a NewsGuard score higher than 60 is often labeled as reliable; otherwise, it is unreliable. NewsGuard has provided ground truth for the construction of news datasets such as NELA-GT-2018~\cite{norregaard2019nela} for studying misinformation.

\paragraph{Media Bias/Fact Check (MBFC)}\footnote{\url{https://mediabiasfactcheck.com/}}
MBFC is a website that rates factual accuracy and political bias of news medium. The fact-checking team consists of Dave Van Zandt, the primary editor and the website owner, and some journalists and researchers (more details can be found on its ``About'' page). MBFC labels each news media as one of six factual-accuracy levels based on the fact-checking results of the news articles it has published (more details can be found on its ``Methodology'' page): 
(i) very high, 
(ii) high,
(iii) most factual,
(iv) mixed,
(v) low, and 
(vi) very low.
Such information has been used as ground truth for automatic fact-checking studies.~\cite{baly2018predicting}

\paragraph{What Are Our Criteria?} Referenced by NewsGuard and MBFC, our criteria for determining reliable and unreliable news sites are:
\begin{itemize}[leftmargin=20mm]
    \item[\color{blue} $\checkmark$ Reliable\hspace{0.25cm}] A news site is reliable if its NewsGuard score is greater than 90, \underline{\textbf{and}} its factual reporting on MBFC is very high or high.
    \item[\color{red} $\times$ Unreliable] A news site is unreliable if its NewsGuard score is less than 30, \underline{\textbf{and}} its factual reporting on MBFC is below mixed.
\end{itemize}

Our search towards news medium with high credibility is conducted among news medium listed in MBFC ($\sim$2,000). To find news medium with low credibility, we search in MBFC and the newly released 
``Coronavirus Misinformation Tracking Center''\textsuperscript{\ref{footnote:newsguard}} of NewsGuard, which provides a list of websites publishing false coronavirus information.
Ultimately, we obtain a total of 60 news sites, from which 22 are the sources of reliable news articles (e.g., \textit{National Public Radio}\footnote{\url{https://www.npr.org}} and \textit{Reuters}\footnote{\url{https://www.reuters.com}}) and the remaining 38 are sources to collect unreliable news articles (e.g., \textit{Human Are Free}\footnote{\url{http://humansarefree.com/}} and \textit{Natural News}\footnote{\url{https://www.naturalnews.com}}). The full list of sites considered in our repository is also available at \url{http://coronavirus-fakenews.com}. Note that several ``fake'' news medium are not included, such as \textit{70 News}, \textit{Conservative 101}, and \textit{Denver Guardian}, since they no longer exist or their domains have been unavailable.

\begin{figure}[t]
\centering
\begin{minipage}{\columnwidth}
    \subfigure[Reliable News Sites]{
    \includegraphics[width=0.49\columnwidth]{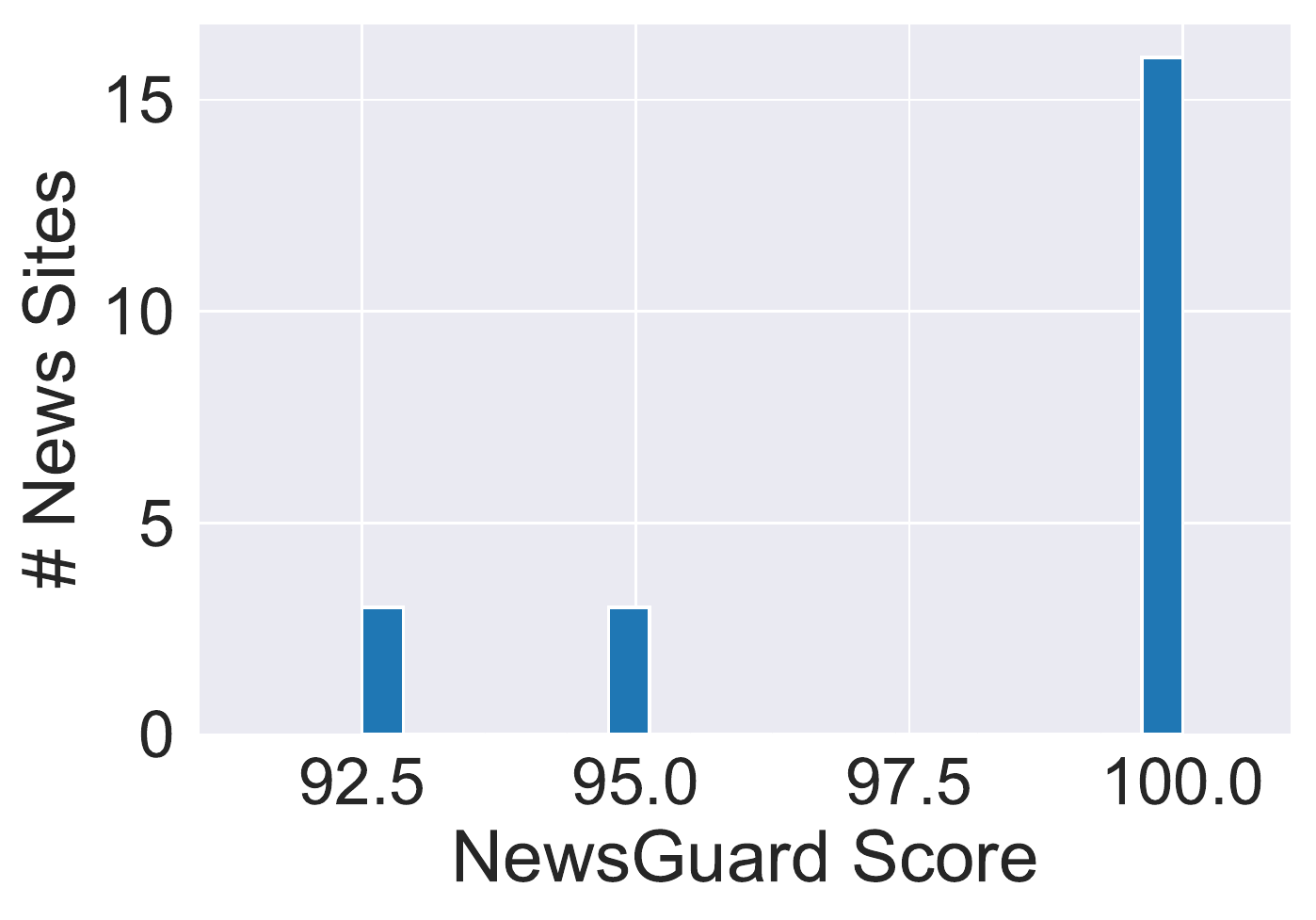}
    \includegraphics[width=0.49\columnwidth]{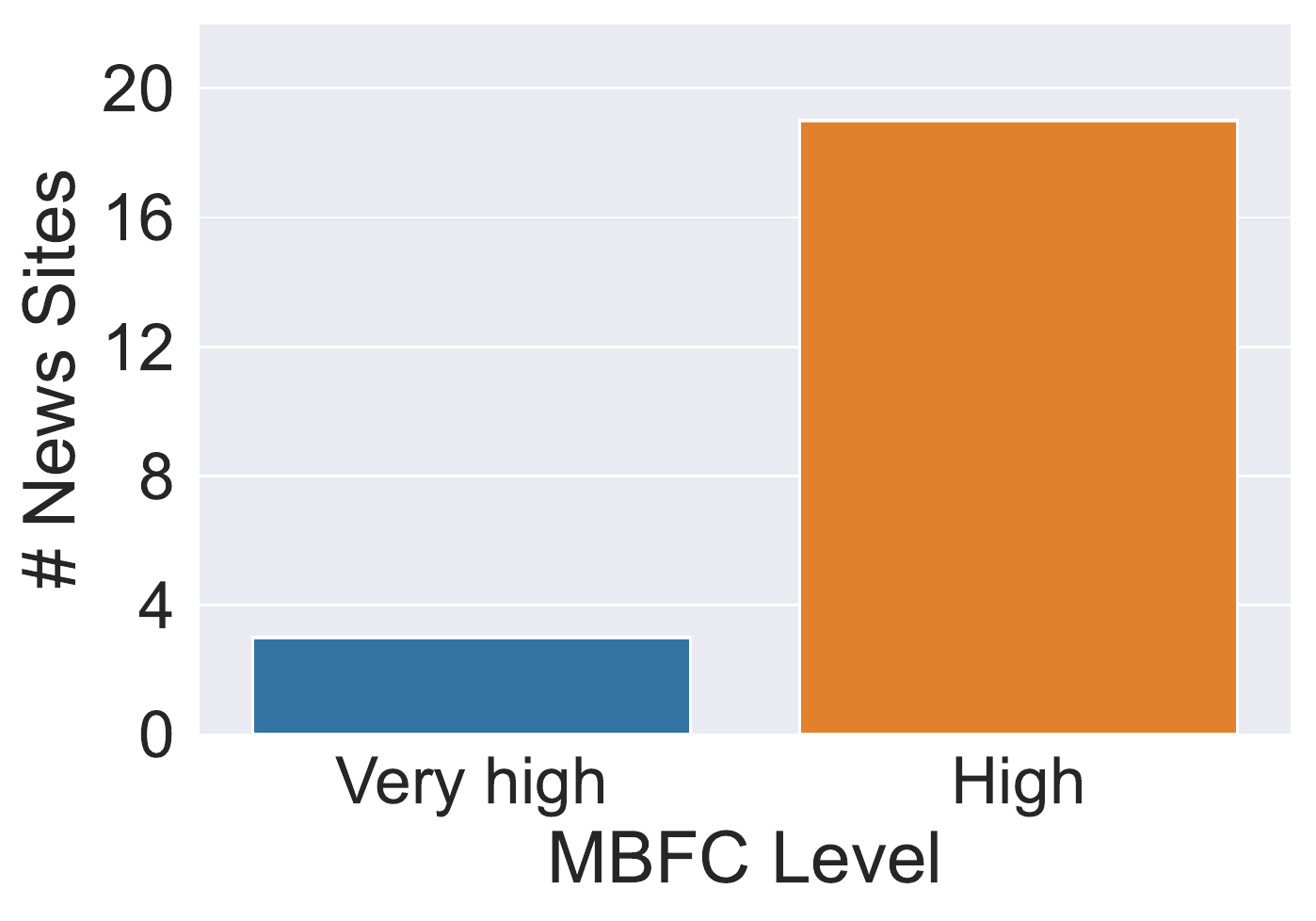}}\quad
    \subfigure[Unreliable News Sites]{
    \includegraphics[width=0.49\columnwidth]{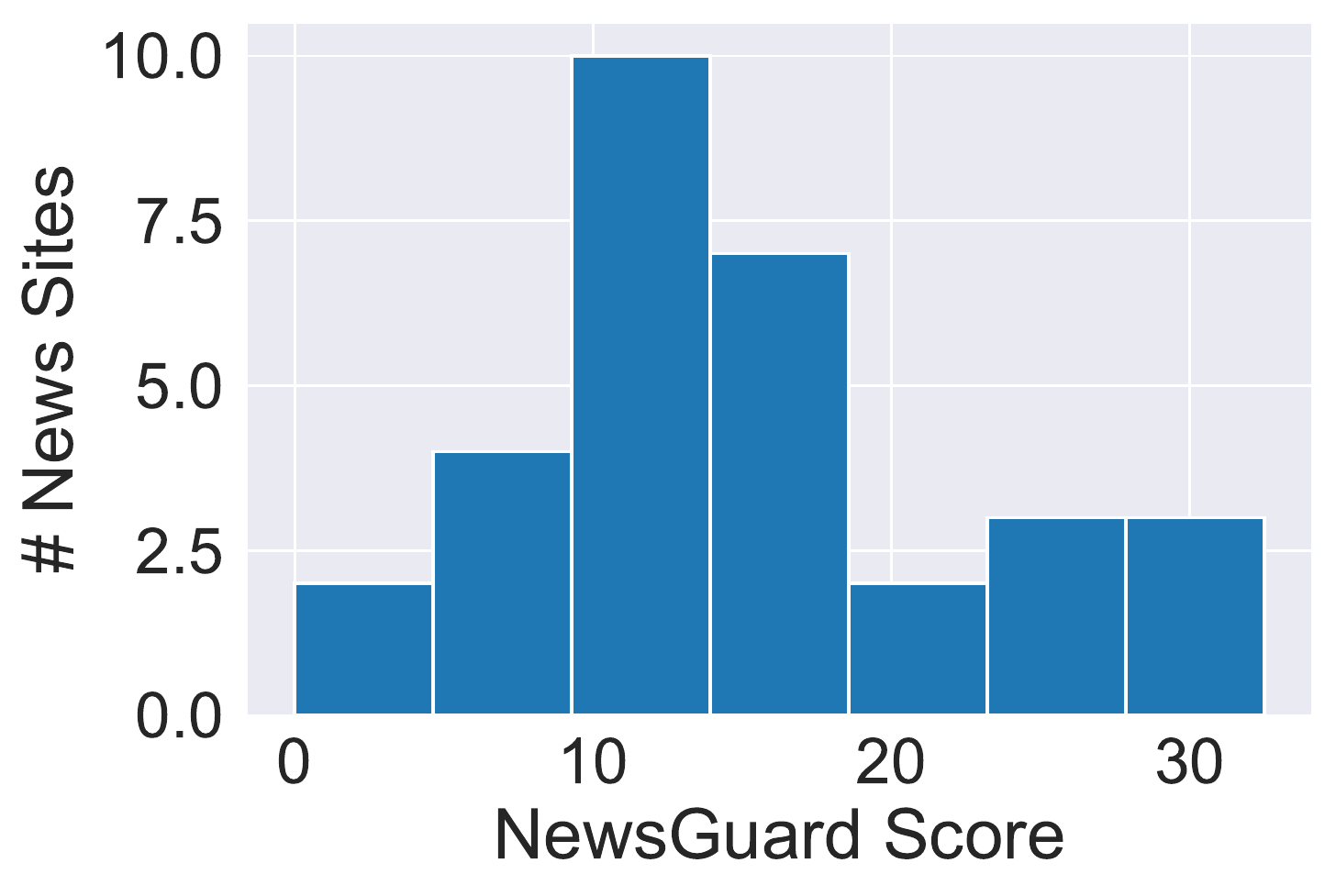}
    \includegraphics[width=0.49\columnwidth]{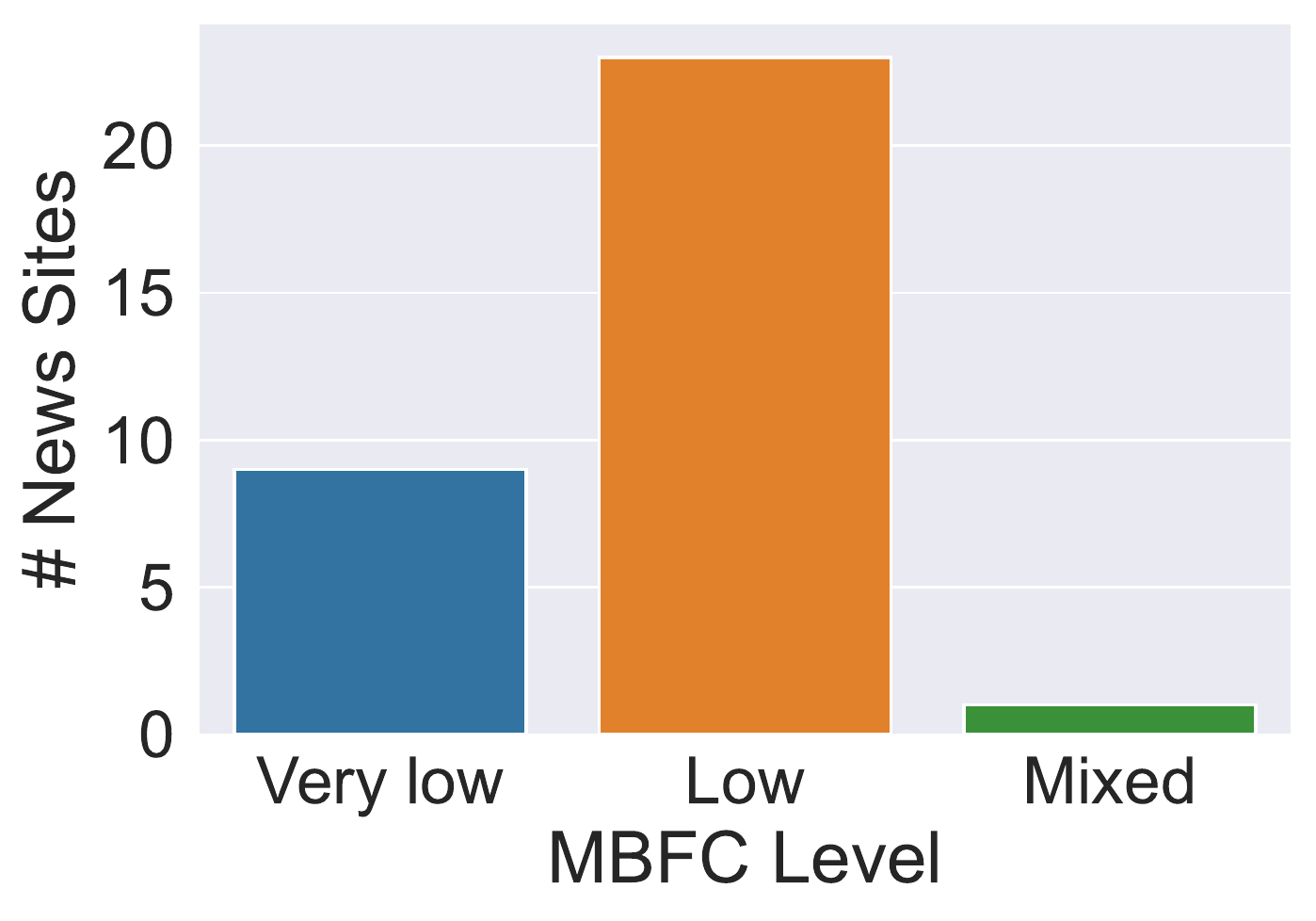}}
\end{minipage}
\caption{Credibility Distribution of Determined News Sites}
\label{fig:site_cred_dist}
\end{figure}

\begin{figure}[t]
    \subfigure[Reliable News\protect\footnotemark]{\label{subfig:cred_news_exp}
    \includegraphics[width=0.46\columnwidth]{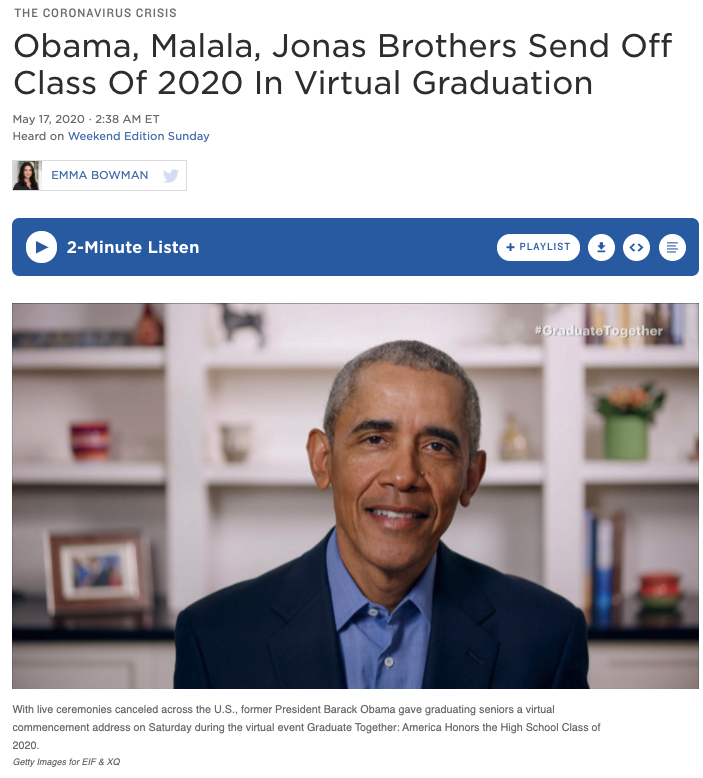}}~
    \subfigure[Unreliable News\protect\footnotemark]{\label{subfig:uncred_news_exp}
    \includegraphics[width=0.5\columnwidth]{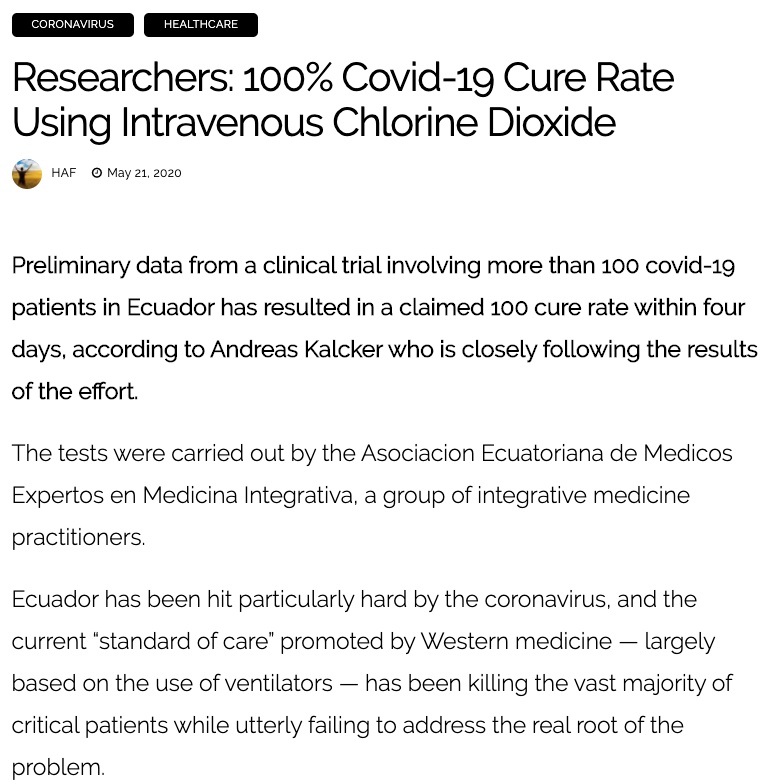}}
    \caption{Examples of News Articles Collected}
    \label{fig:news_examples}
\end{figure}

\footnotetext[15]{\url{https://www.npr.org/sections/coronavirus-live-updates/2020/05/17/857512288/obama-malala-jonas-brothers-send-off-class-of-2020-in-virtual-graduation}}

\footnotetext[16]{\url{https://humansarefree.com/2020/05/researchers-100-covid-19-cure-rate-using-intravenous-chlorine-dioxide.html}}

Also note that to achieve a good trade-off between dataset scalability and label accuracy, we utilize more extreme threshold scores (30 and 90) compared to the initial one provided by NewsGuard (60). In this way, the selected news sites exhibit an \textit{extreme} reliability (or unreliability), which helps reduce the number of false positives and false negatives in news labels in our repository; ideally, each news article published on a reliable site is factual, and on an unreliable site is false. 
Figure \ref{fig:site_cred_dist} illustrates the 
credibility distributions of reliable and unreliable news sites. It can be observed from the figure that for reliable news, most of them have a full mark on NewsGuard and are labeled as ``high"ly factual by MBFC; ``very high'' is rare for all sites listed in MBFC. In contrast, unreliable news sites share an average NewsGuard score of $\sim$15 and a low factual label by MBFC; similarly, ``very low'' is rarely given on MBFC.

\subsection{Collecting COVID-19 News Content}
\label{subsec:news_content}

To crawl COVID-19 news articles from selected news sites, we first determine whether the news article is about COVID-19; the process is detailed in Section~\ref{subsubsec:keywords}. Next, we detail how the data is crawled and the news content components that are included in our repository in Section \ref{subsubsec:components}.

\subsubsection{News Topic Identification} 
\label{subsubsec:keywords}

To identify news articles on COVID-19, we use a list of keywords: 
\begin{itemize}
    \item \texttt{SARS-CoV-2},
    \item \texttt{COVID-19}, and 
    \item \texttt{Coronavirus}.
\end{itemize}
News articles whose content contains any of the keywords (case-insensitive) are considered related to COVID-19. These three keywords are the official names announced by the WHO on February $11^{\text{th}}$, where SARS-CoV-2 (standing for Severe Acute Respiratory Syndrome CoronaVirus 2) is the virus name, and Coronavirus and COVID-19 are the name of the disease that the virus causes. Before the WHO announcement, COVID-19 was previously known as the ``2019 novel coronavirus,''\footnote{\url{https://www.who.int/emergencies/diseases/novel-coronavirus-2019/technical-guidance/naming-the-coronavirus-disease-(covid-2019)-and-the-virus-that-causes-it}\label{footnote:naming}}, which also includes the \texttt{coronavirus} keyword which we are considering.  We merely consider official names as keywords to avoid potential biases or even discrimination in naming. Furthermore, a news media (article) that is credible, or pretends to be credible, often acts professionally and adopts the official name(s) of the disease/virus. Compared to those articles that use biased and/or inaccurate terms, false news pretending to be professional is more detrimental and challenging to detect, which has become the focus of current fake news studies.~\cite{zhou2020survey} Examples of such news articles are illustrated in Figure \ref{fig:news_examples}. 

\subsubsection{Crawling News Content}
\label{subsubsec:components}

Content crawler relies on \textsf{Newspaper} Python library.\footnote{\url{https://github.com/codelucas/newspaper}} The content of each news article corresponds to twelve components:
\begin{enumerate}
    \item[(\textbf{C1})] \textit{News ID}: Each news article is assigned a unique id as the identity;
    \item[(\textbf{C2})] \textit{News URL}: The URL of the news article. The URL helps us verify the correctness of the collected data. It can also be used as the reference and source when repository users would like to extend the repository by fetching additional information;
    \item[(\textbf{C3})] \textit{Publisher}: The name of the news media (site) that publishes the news article;
    \item[(\textbf{C4})] \textit{Publication Date}: The date (in \texttt{yyyy-mm-dd} format) on which the news article was published on the site, which provides temporal information to support the investigation of, e.g., the relationship between the misinformation volume and the outbreak of COVID-19 over time;
    \item[(\textbf{C5})] \textit{Author}: The author(s) of the news article, whose number can be none, one, or more than one. Note that some news articles might have fictional author names. Author information is valuable in evaluating news credibility by either investigating the collaboration network of authors~\cite{sitaula2020credibility} or exploring its relationships with news publishers and content~\cite{zhang2020fakedetector};
    \item[(\textbf{C6-7})] \textit{News Title and Bodytext} as the main textual information; 
    \item[(\textbf{C8})] \textit{News Image} as the main visual information, which is provided in the form of a link (URL). Note that most images within the news page are noise -- they can be advertisements, images belonging to other news articles due to the recommender systems embedded in news sites, logos of news sites and/or social media icons, such as Twitter and Facebook logos for sharing. Hence, we particularly fetch the main/head/top image for each news article to reduce noise; 
    \item[(\textbf{C9})] \textit{Country}: The name of country where the news is published;
    \item[(\textbf{C10})] \textit{Political bias}: Each news article is labeled as one of `extremely left', `left', `left-center', `center', `right-center', `right', and `extremely right' that is equivalent to the political bias of its publisher. News political bias is verified by two resources, AllSides\footnote{\url{https://www.allsides.com/unbiased-balanced-news}} and MFBC, both of which rely on domain experts to label media bias; and 
    \item[(\textbf{C11-12})] \textit{NewsGuard score and MBFC factual reporting} as the original ground truth of news credibility, which has been detailed in Section \ref{subsec:news_sites}. 
\end{enumerate}

\subsection{Tracking News Spreading on Social Media}
\label{subsec:social_media}

We first use Twitter Premium Search API\footnote{\url{https://developer.twitter.com/en/docs/tweets/search/overview/premium}} to track the spread of collected news articles on Twitter. Specifically, our search is based on the URL of each news article and looks for tweets posted after the date when the news article was published to the current date (for the current version of the dataset, this date is May $26^{\text{th}}$). Twitter Search API can return the corresponding tweets with detailed information such as their IDs, text, languages of text, times of being created, statistics on retweeted/replied/liked. Also, it returns the information of users who post these tweets, such as user IDs and their number of followers/friends. To comply with Twitter's Terms of Service,\footnote{\url{https://developer.twitter.com/en/developer-terms/agreement-and-policy}} we only publicly release the IDs of the collected data for non-commercial research use, but provide the instructions for obtaining the tweets using the released IDs for user convenience. More details can be seen in \url{http://coronavirus-fakenews.com}.

\section{Data Statistics and Distributions}
\label{sec:data_analysis}

The general statistics on our dataset is presented in Table \ref{tab:data_statistics}. The dataset contains 2,029 news articles, most of which have both textual and visual information for multimodal studies (\#=2,017),~\cite{zhou2020multimodal} and have been shared on social media (\#=1,747). The dataset is imbalanced in news class -- the proportion of reliable versus unreliable news articles is around 2:1. The number of users who spread reliable news (\#=78,659) plus that of users spreading unreliable news (\#=17,323) is greater than the total number of users included in the dataset (\#=93,761). This observation indicates that users can both engage in spreading reliable and unreliable news articles.

\begin{table}
\centering
\caption{Data Statistics}
\label{tab:data_statistics}
\begin{tabular}{lrrr}
\toprule[1pt]
 & \textbf{Reliable} & \textbf{Unreliable} & \textbf{Total} \\ \midrule[0.5pt]
\textbf{News articles} & 1,364 & 665 & 2,029 \\
\quad w/ images & 1,354 & 663 & 2,017 \\
\quad w/ social information & 1,219 & 528 & 1,747 \\
\textbf{Tweets} & 114,402 & 26,418 & 140,820 \\
\textbf{Users} & 78,659 & 17,323 & 93,761 \\
\bottomrule[1pt]
\end{tabular}
\end{table}

\begin{figure*}[t]
    \includegraphics[width=.94\textwidth]{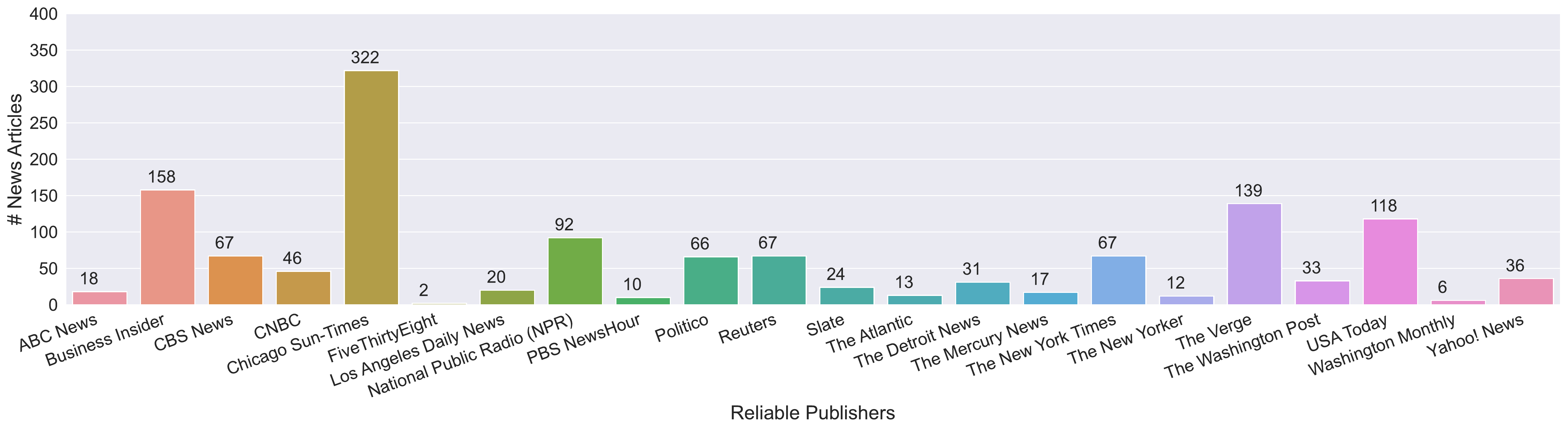}
    \includegraphics[width=.94\textwidth]{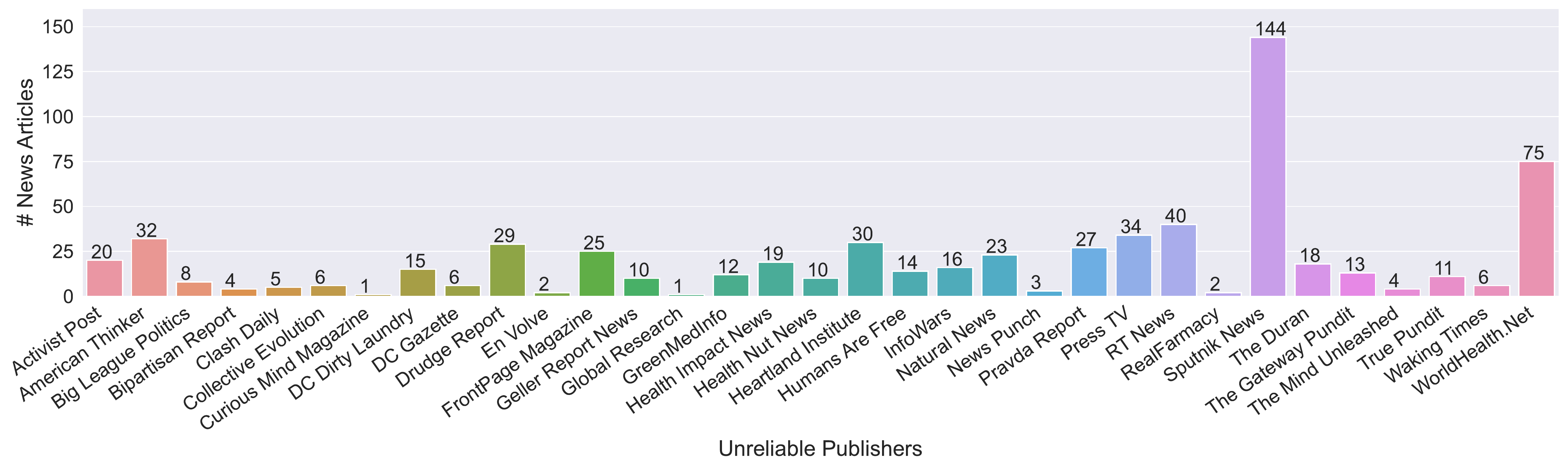}
    \caption{Distribution of News Publishers}
    \label{fig:publisher_dist}
\end{figure*}

\begin{figure}[t]
    \begin{minipage}{\columnwidth}
    \begin{minipage}{0.49\textwidth}
    \centering
    \includegraphics[width=\textwidth]{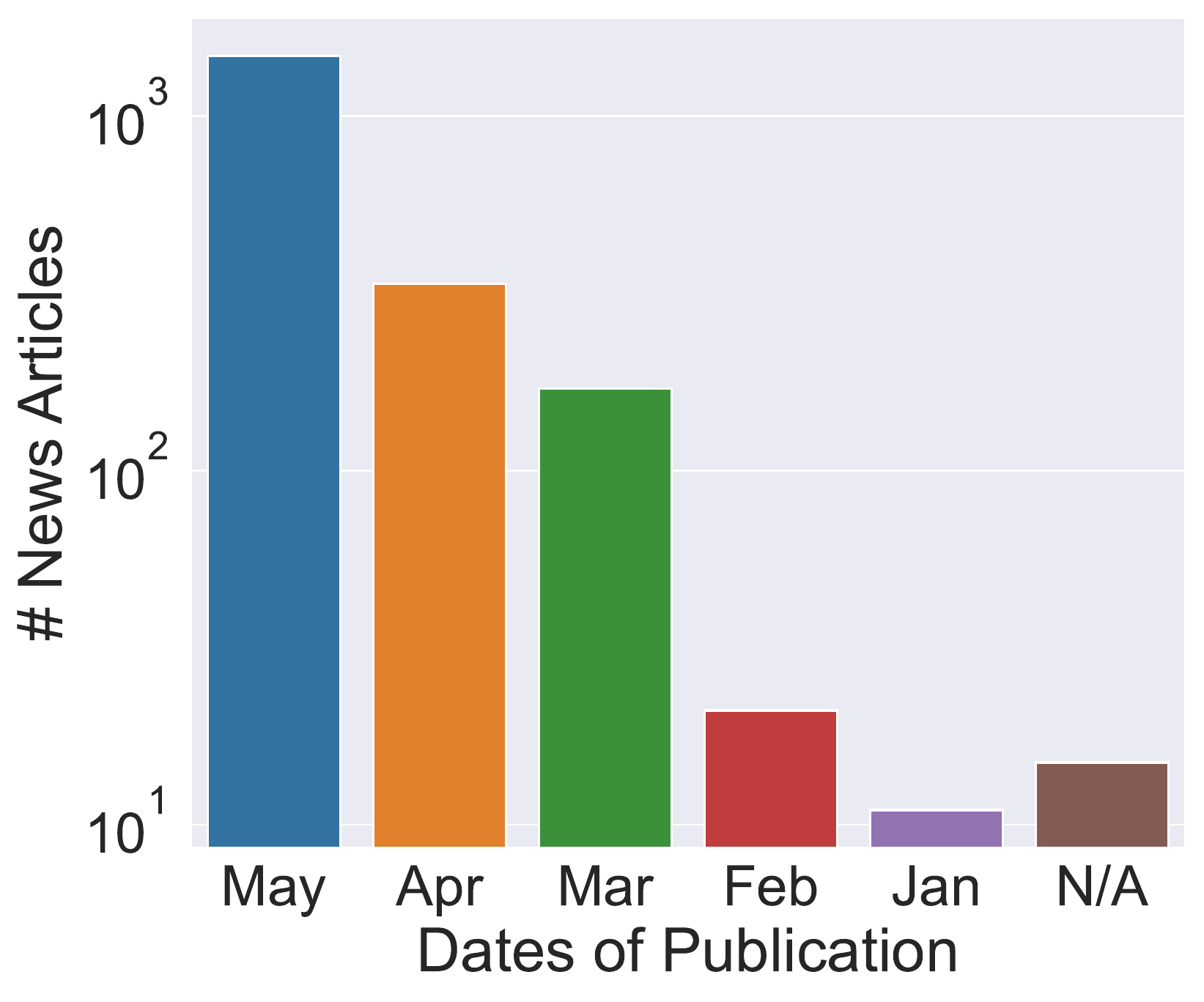}
    \caption{Publication Date}
    \label{fig:publish_date_dist}
    \end{minipage}
    \begin{minipage}{0.49\textwidth}
    \centering
    \includegraphics[width=\textwidth]{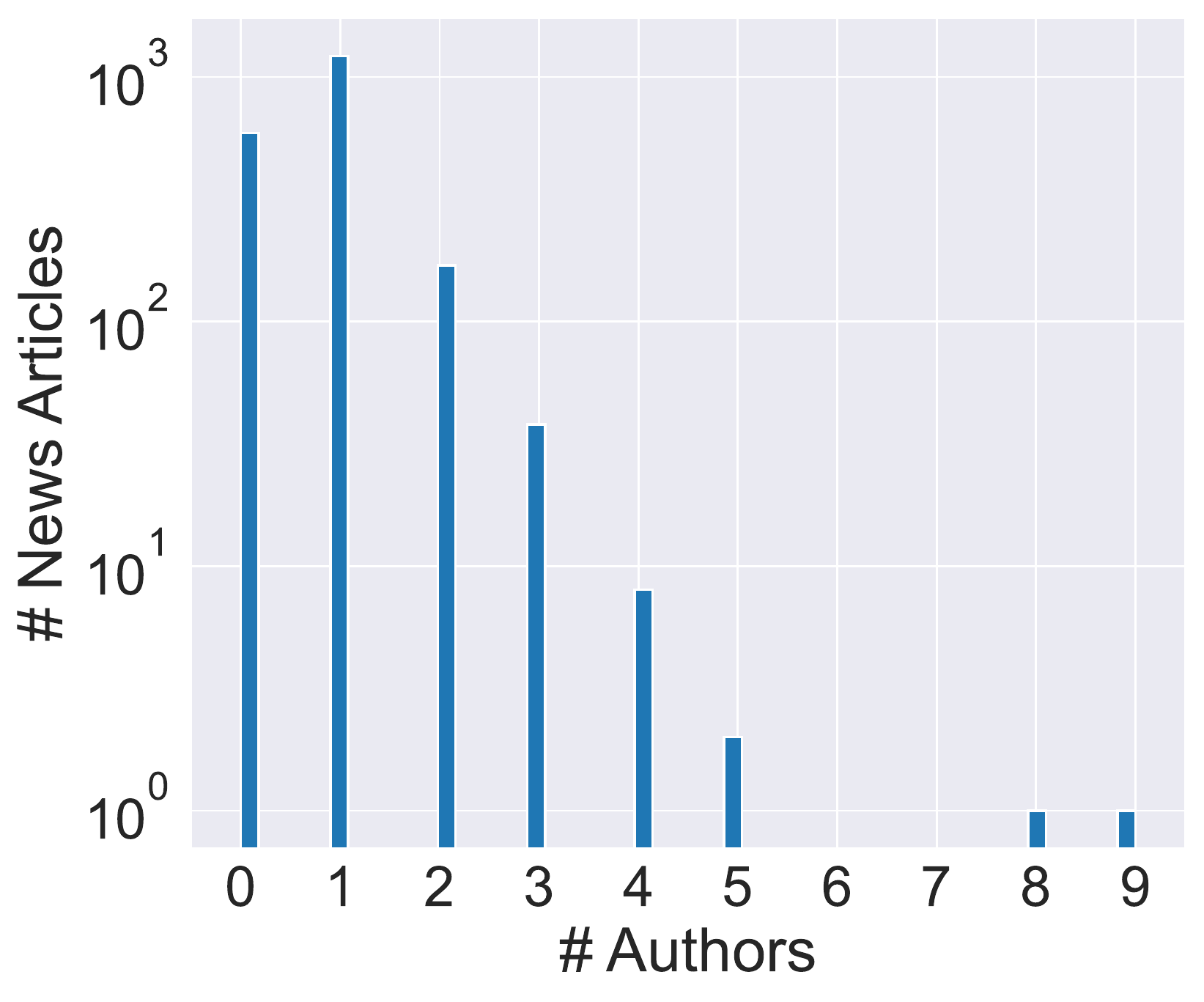}
    \caption{Author Count}
    \label{fig:author_dist}  
    \end{minipage}
    \end{minipage}
\end{figure}

Next, we visualize the distributions of data features/attributes.

\vspace{-1mm}
\paragraph{Distribution of News Publishers}
Figure \ref{fig:publisher_dist} shows the number of COVID-19 news articles published in each [extremely reliable or extremely unreliable] news site. There are five unreliable publishers with no news on COVID-19; hence, they are not presented in the figure. We keep these publishers in our repository as the data will be updated over time and these publishers may publish news articles on COVID-19 in the future.

\vspace{-1mm}
\paragraph{News Publication Dates}
The distribution of news publication dates is presented in Figure \ref{fig:publish_date_dist}, where all articles are published in 2020. We point out that from January to May, the number of COVID-19 news articles published is significantly (exponentially) increased. The possible explanation for this phenomena is three-fold.
First, from the time that the outbreak was first identified in Wuhan, China (December 2019)~\cite{huang2020clinical} to May 2020, the number of confirmed cases and deaths caused by SARS-CoV-2 have exponentially grown  globally.\textsuperscript{\ref{footnote:who_report}} Meanwhile, the virus has become a world topic and has triggered more and more discussions on a world-wide scale.
Second, some older news articles are no longer available, which has motivated us to timely update the dataset.
Third, the keywords we have used to identify COVID-19 news articles are the official ones provided by the WHO in February.\textsuperscript{\ref{footnote:naming}} Some news articles published in January are also collected, as before the WHO announcement COVID-19 was known as the ``2019 novel coronavirus,'' which also includes one of our keywords ``coronavirus.'' We have detailed the reasons behind our keyword selection in Section \ref{subsubsec:keywords}. Note that there are a small group of news articles whose publication dates are not accessible, which we denote as N/A in Figure \ref{fig:publish_date_dist}.

\begin{figure}[t]
    \subfigure[Network]{
    \includegraphics[width=0.44\columnwidth]{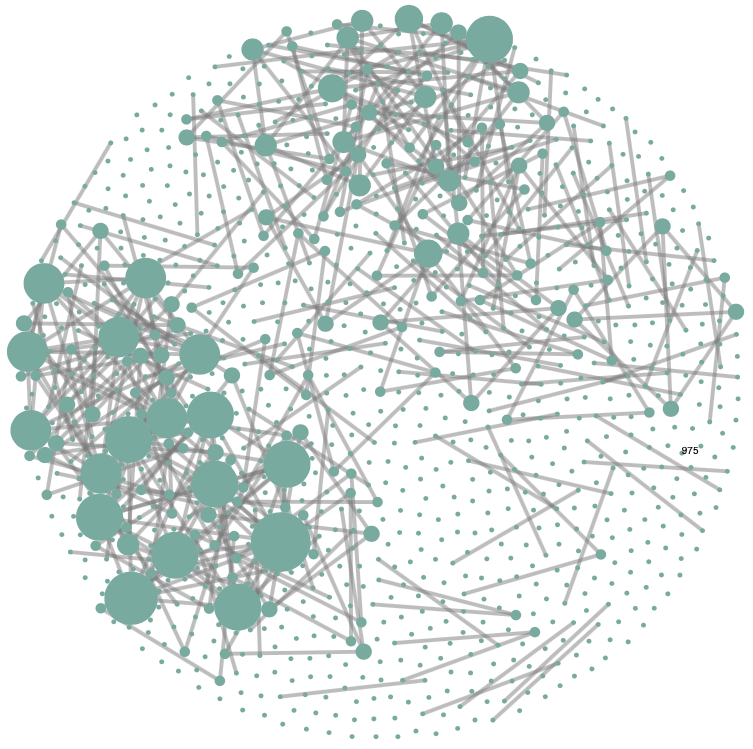}}~
    \subfigure[Degree Distribution]{
    \includegraphics[width=0.52\columnwidth]{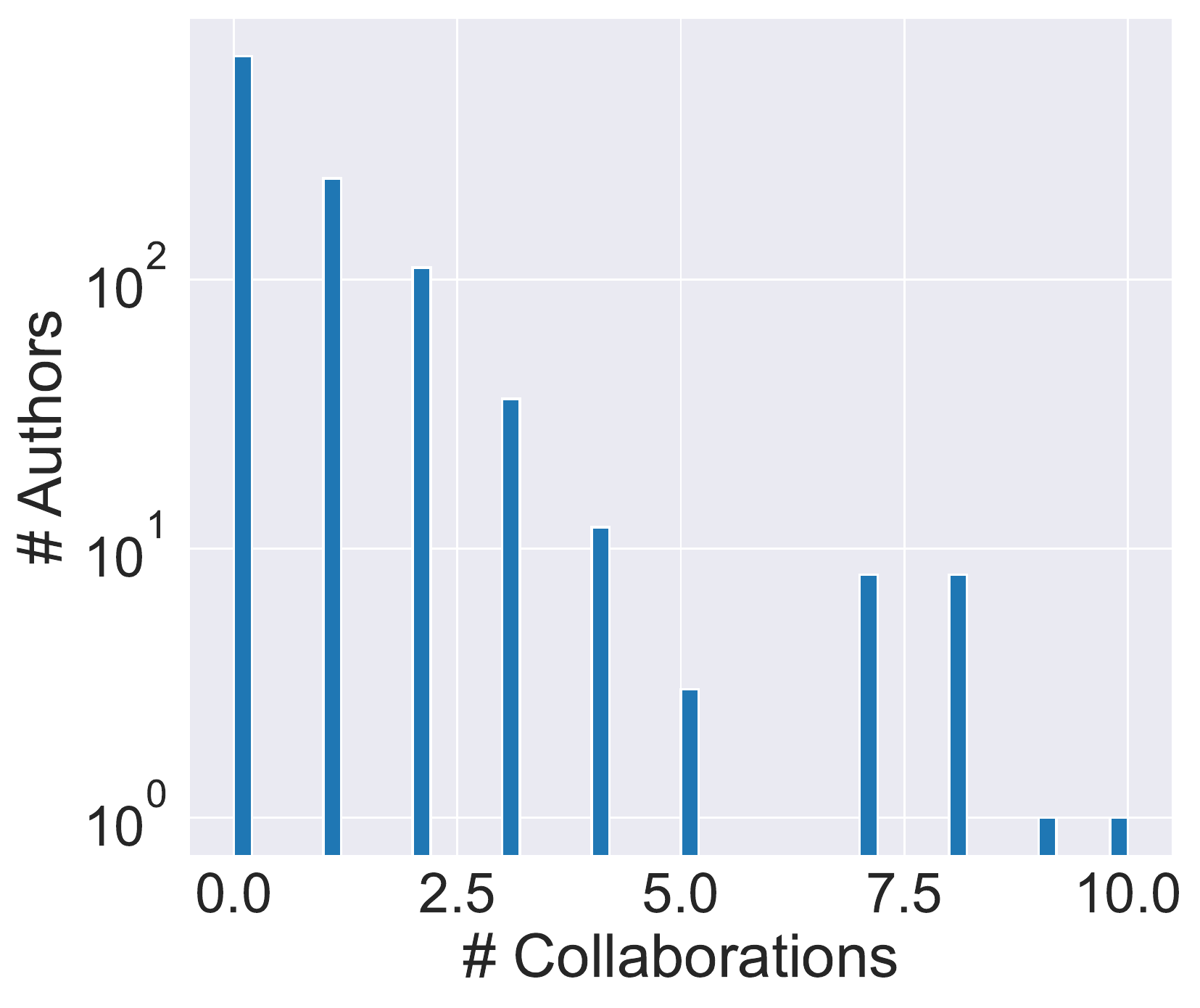}}
    \vspace{-2mm}
    \caption{Author Collaborations}
    \label{fig:author_collaboration}
    \vspace{3mm}
    \begin{minipage}{\columnwidth}
    \begin{minipage}{0.54\columnwidth}
    \centering
    \includegraphics[width=\columnwidth]{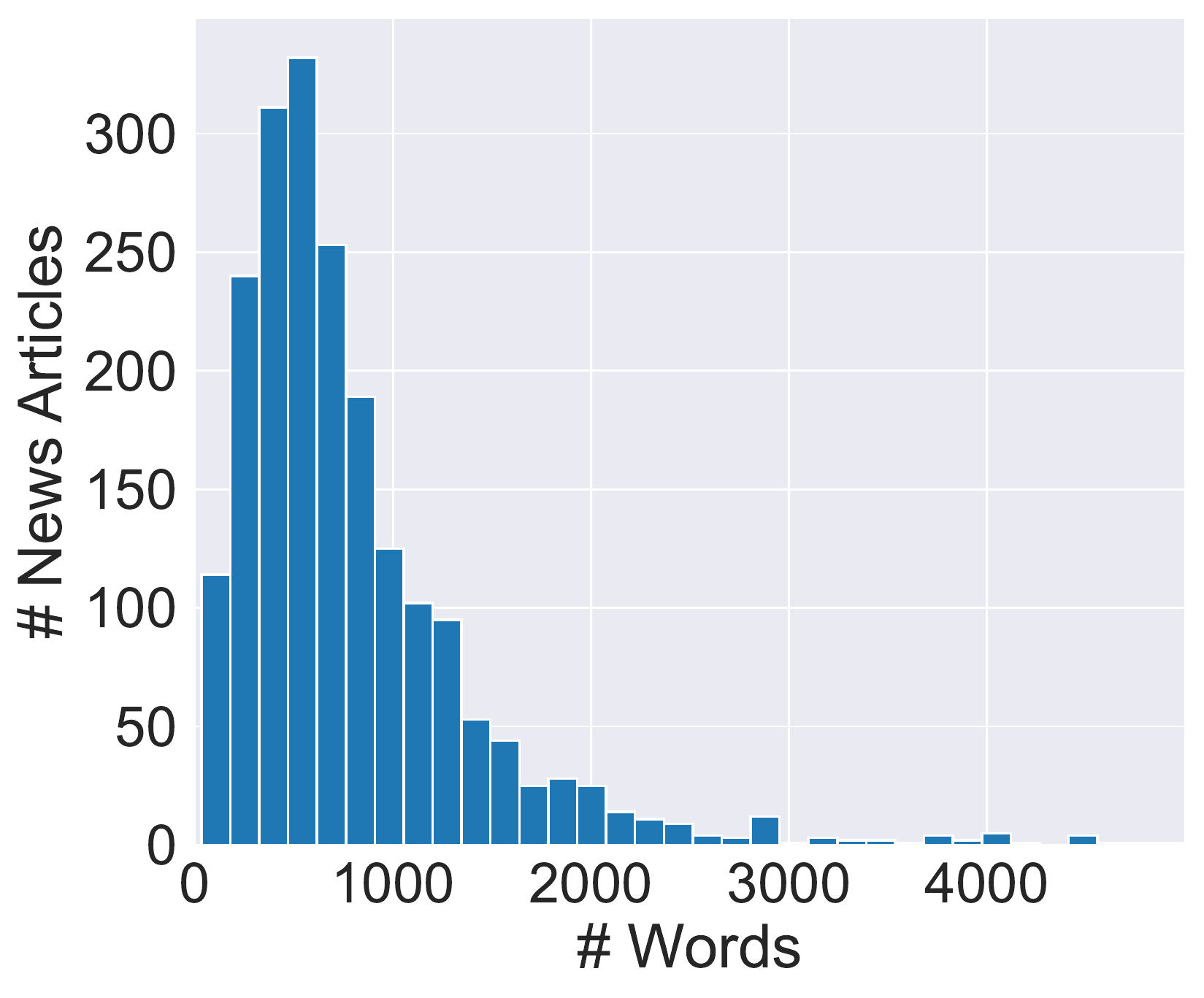}
    \caption{Word Count}
    \label{fig:word_count}
    \end{minipage}
    \begin{minipage}{0.45\columnwidth}
    \centering
    \includegraphics[width=\columnwidth]{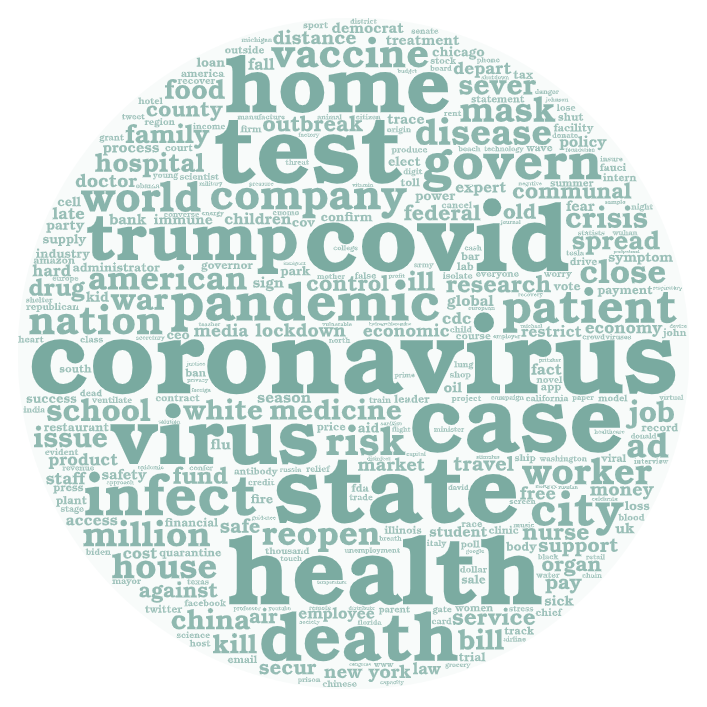}
    \caption{Word Cloud}
    \label{fig:word_cloud} 
    \end{minipage}
    \end{minipage}
\end{figure}

\vspace{-1mm}
\paragraph{News Authors and Author Collaborations}
Figure \ref{fig:author_dist} presents the distribution of the number of authors contributing to news articles, which is governed by a long-tail distribution: most articles are contributed by $\leq 5$ authors. Instead of including the [real or virtual] names of the authors, some articles provide publisher names as authors. Considering such information has been available in the repository, we leave the author information of these news articles blank, i.e., their number of authors is zero. Furthermore, we construct the coauthorship network, shown in Figure \ref{fig:author_collaboration}. It can be observed from the network that node degrees also follow a power-law-like distribution: among 1,095 nodes (authors), over 90\% of them have less than or equal to two collaborators. 

\vspace{-1mm}
\paragraph{News Content Statistics}
Both Figures \ref{fig:word_count} and \ref{fig:word_cloud} reveal textual characteristics within news content (including news title and bodytext). It can be observed from Figure \ref{fig:word_count} that the number of words within news content follows a long-tail (power-low-like) distribution, with an average value of $\sim$800 and a median value of $\sim$600.
On the other hand, Figure \ref{fig:word_cloud} provides the word cloud for the entire repository. As the news articles collected share the same COVID-19 topic, some relevant topics and vocabularies have been naturally and frequently used by the news authors, such as ``coronavirus'' (\#=6465),  ``COVID'' (\#=5413), ``state'' (\#=4432), ``test'' (\#=4274), ``health'' (\#=3714), ``pandemic'' (\#=3427), ``virus'' (\#=2903), ``home'' (\#=2871), ``case'' (\#=2676), and ``Trump'' (\#=2431) that are illustrated with word font size scaled to their frequencies. 

\begin{figure*}[t]
\begin{minipage}{\columnwidth}
    \subfigure[News Publishers]{
    \includegraphics[width=0.48\columnwidth]{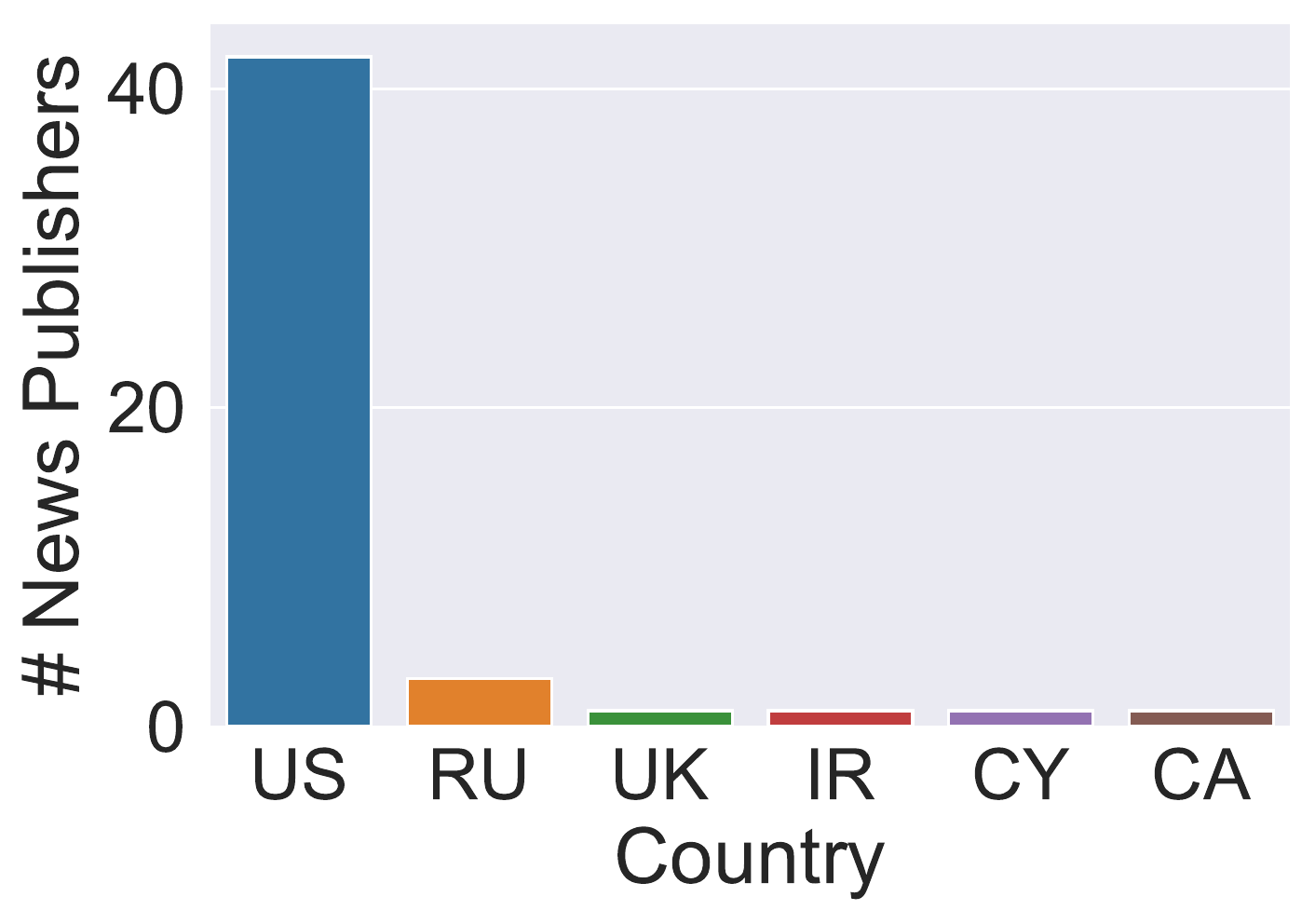}}
    \subfigure[News Articles]{
    \includegraphics[width=0.5\columnwidth]{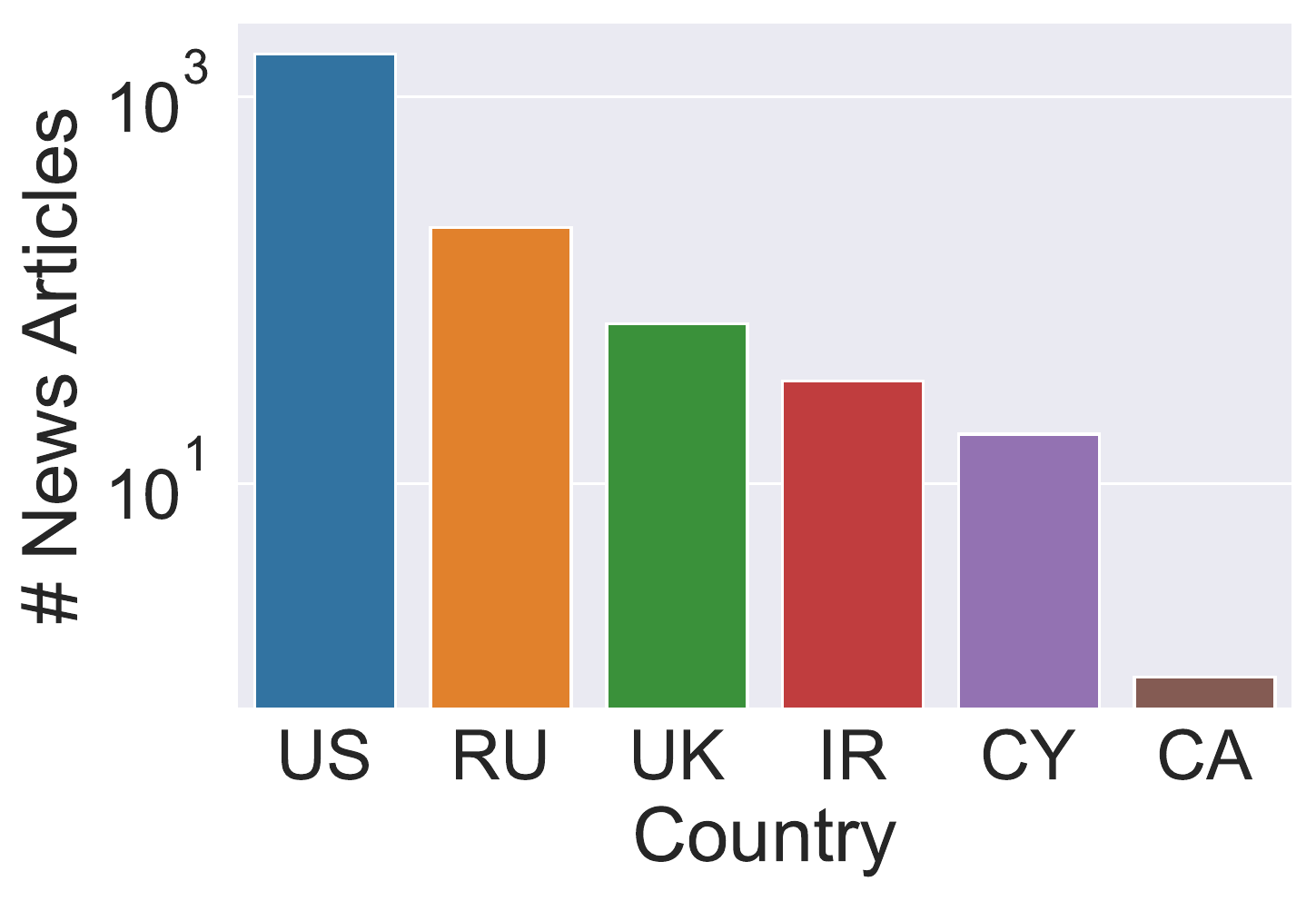}}  
    \vspace{-3mm}
    \caption{Country}
    \label{fig:country_dist}
\end{minipage}
\begin{minipage}{\columnwidth}
    \subfigure[News Publishers]{
    \includegraphics[width=0.48\columnwidth]{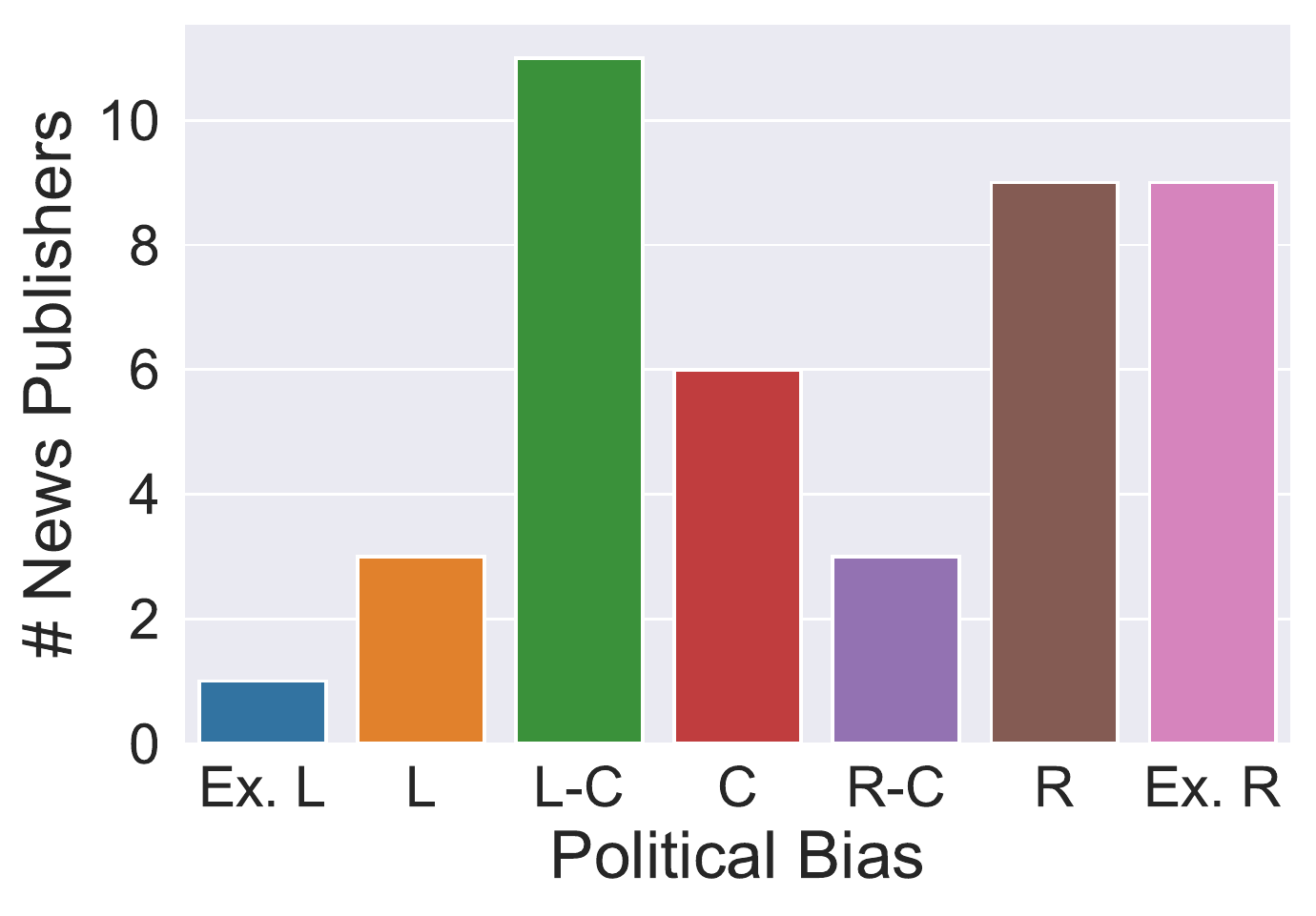}}
    \subfigure[News Articles]{
    \includegraphics[width=0.5\columnwidth]{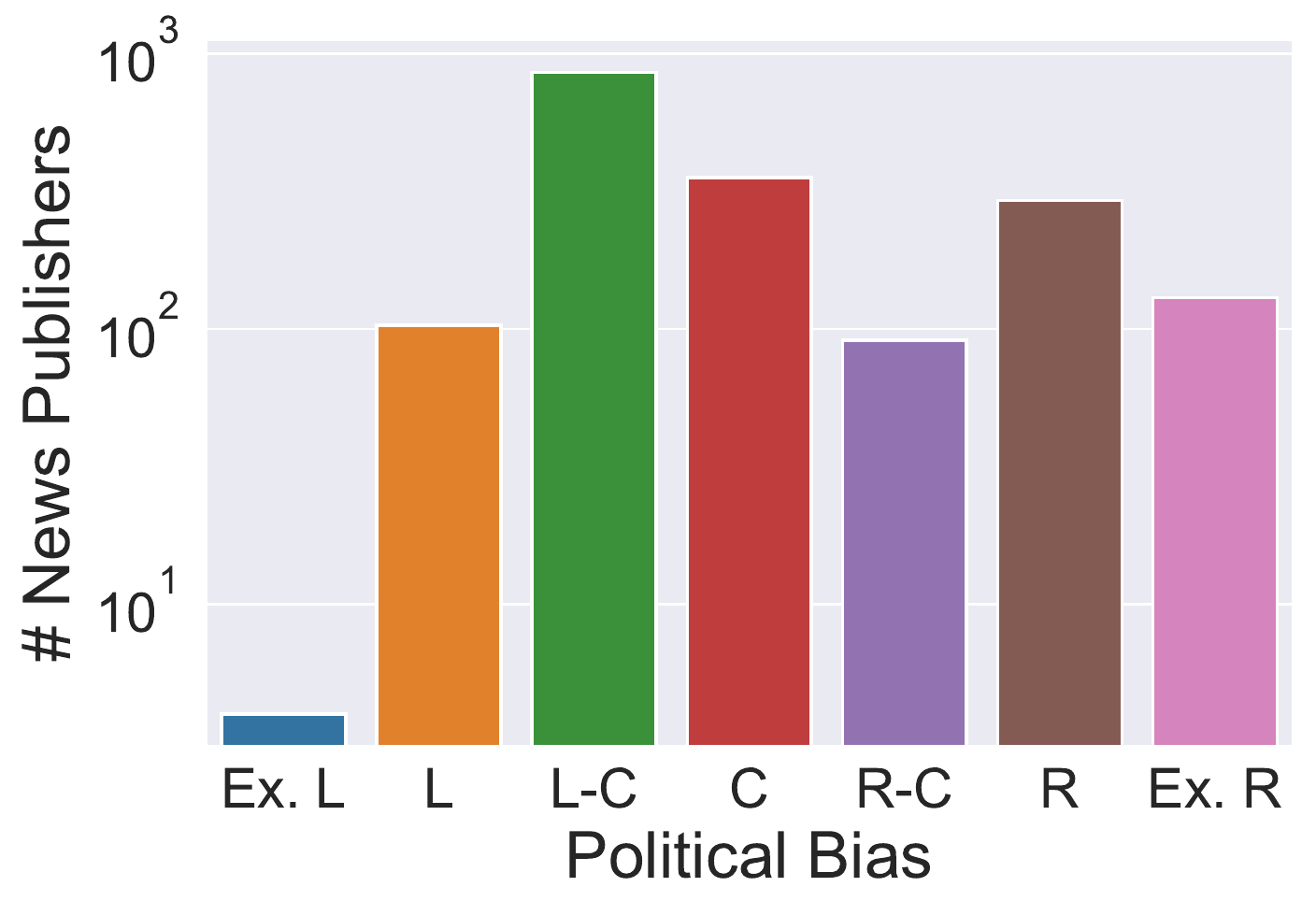}}  
    \vspace{-3mm}
    \caption{Political Bias}
    \label{fig:bias_dist}
\end{minipage}
\end{figure*}

\begin{figure*}[t]
    \begin{minipage}{\columnwidth}
    \begin{minipage}{0.5\columnwidth}
    \centering
    \includegraphics[width=\columnwidth]{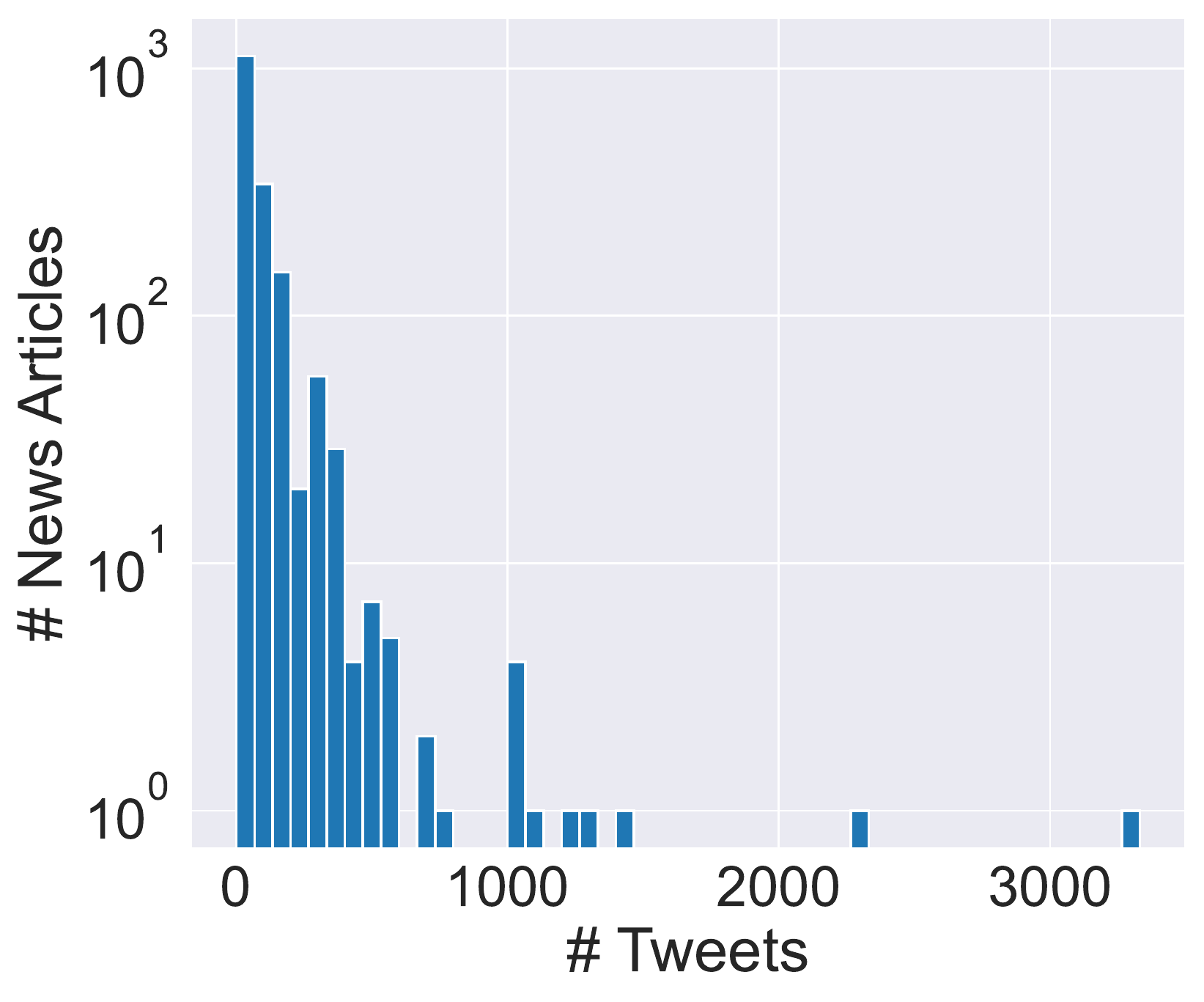}
    \vspace{-3mm}
    \caption{\small{Spreading Frequency}}
    \label{fig:tweet_dist}
    \end{minipage}
    \begin{minipage}{0.5\columnwidth}
    \centering
    \includegraphics[width=\columnwidth]{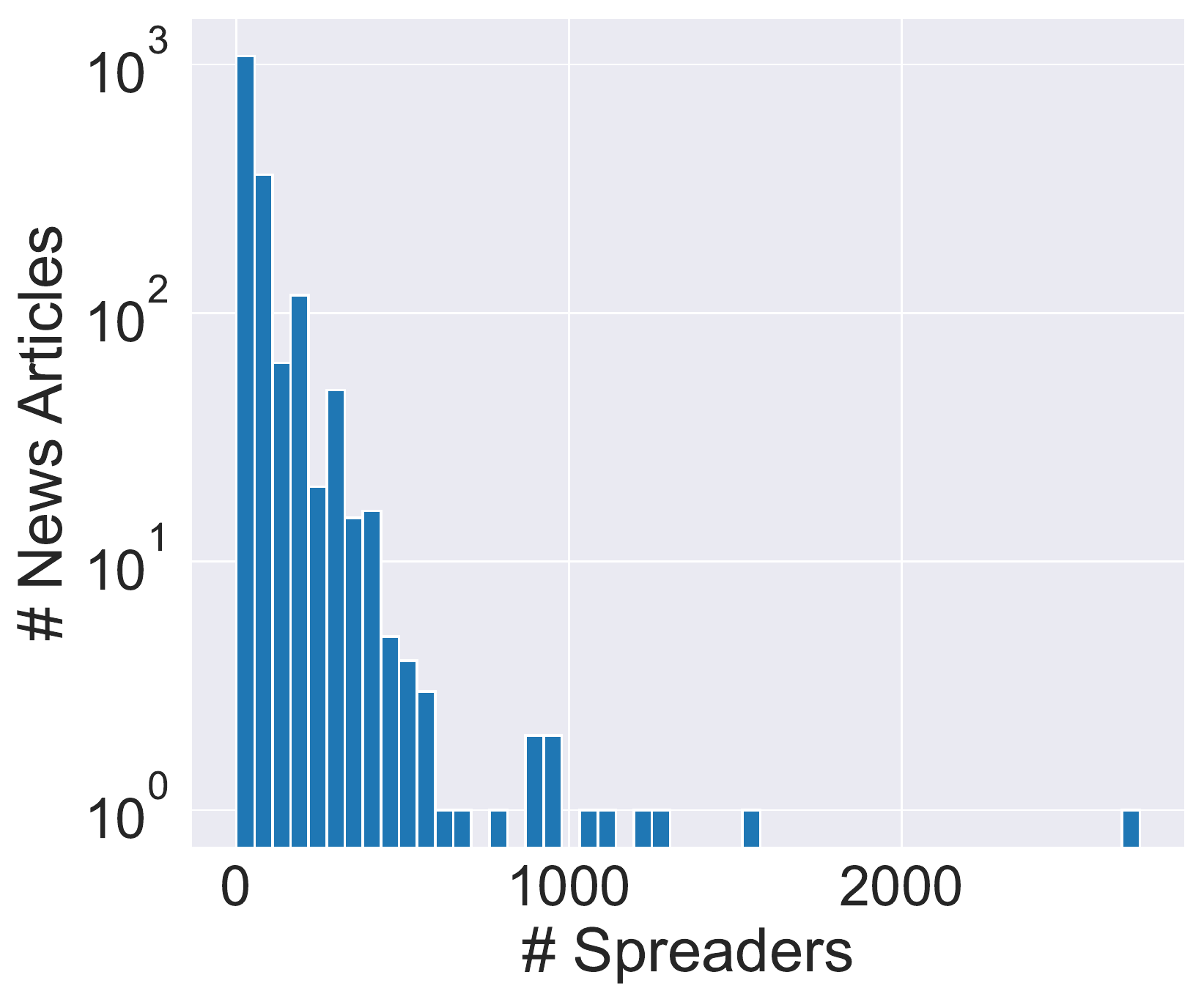}
    \vspace{-3mm}
    \caption{\small{News Spreaders}}
    \label{fig:spreader_dist} 
    \end{minipage}
    \end{minipage}
    \begin{minipage}{\columnwidth}
    \begin{minipage}{0.5\columnwidth}
    \centering
    \includegraphics[width=\columnwidth]{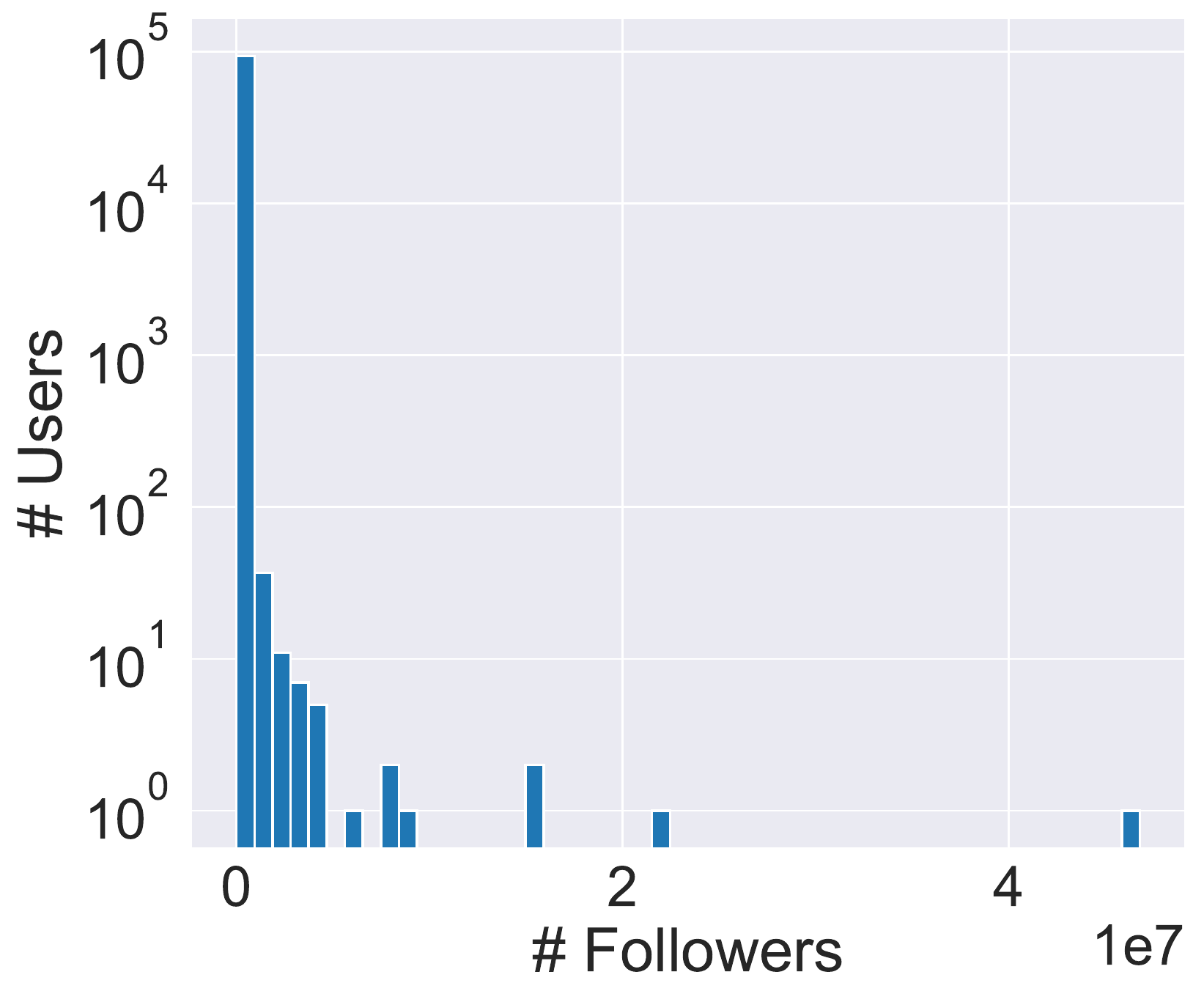}
    \vspace{-3mm}
    \caption{\small{Follower Distribution}}
    \label{fig:follower_dist}
    \end{minipage}
    \begin{minipage}{0.5\columnwidth}
    \centering
    \includegraphics[width=\columnwidth]{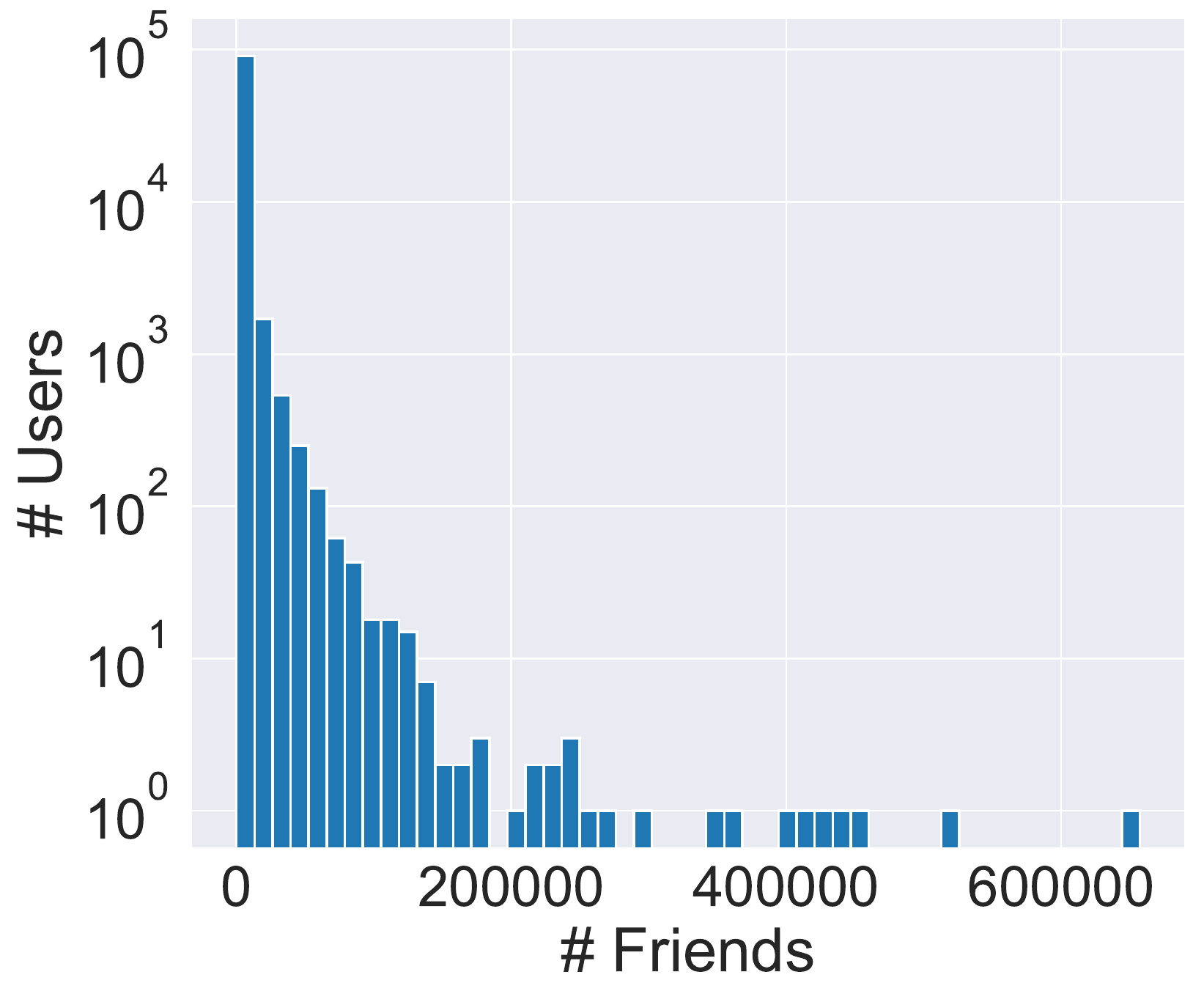}
    \vspace{-3mm}
    \caption{\small{Friend Distribution}}
    \label{fig:friend_dist} 
    \end{minipage}
    \end{minipage}
\end{figure*}

\vspace{-1mm}
\paragraph{Country Distribution}
Figure \ref{fig:country_dist} reveals the countries that news and news publishers belong to. It can be observed that in total six countries -- United States (abbr. US), Russia (abbr. RU), United Kingdom (abbr. UK), Iran (abbr. IR), Cyprus (abbr. CY), and Canada (abbr. CA) -- are covered, where US news and news publishers constitute the majority of the population.

\vspace{-1mm}
\paragraph{Political Bias}
Figure \ref{fig:bias_dist} provides the distribution of political bias of news and news medium (publishers). 
It can be observed from the figure that for both news and publishers, the distribution for those exhibiting a right bias (including extremely right (abbr. Ex. R), right (abbr. R), and right-center (abbr. R-C)) is more balanced compared to those exhibiting a left bias (including extremely left (abbr. Ex. L), left (abbr. L), and left-center (abbr. L-C)).

\vspace{-1mm}
\paragraph{News Spreading Frequencies}
Figure \ref{fig:tweet_dist} shows the distribution of the number of tweets sharing each news article. The distribution exhibits a long tail -- over 80\% of news articles are spread less than 100 times while a few have been shared by thousands of tweets.

\vspace{-1mm}
\paragraph{News Spreaders}
The distribution of the number of spreaders for each news article is shown in Figure \ref{fig:spreader_dist}. It differs from the distribution in Figure \ref{fig:tweet_dist} as one user can spread a news article multiple times. As for social connections of news spreaders, the distributions of their followers and friends are respectively presented in Figures \ref{fig:follower_dist} and \ref{fig:friend_dist}, where the most popular spreader has over 40 million followers (or 600,000 friends).


\section{Forming Baselines: Using $\mathsf{ReCOVery}$ to Predict COVID-19 News Credibility}
\label{sec:experiments}

In this section, several methods that often act as baselines are utilized and developed to predict COVID-19 news credibility using $\mathsf{ReCOVery}$ data, hoping to facilitate future studies. These methods (baselines) are first specified in Section \ref{subsec:baselines}. The implementation details of experiments are then provided in Section \ref{subsec:implementaion_details}.
Finally, we present the performance results for these methods in Section \ref{subsec:performance}.

\subsection{Methods}
\label{subsec:baselines}

We involve the following methods as our as baselines. These methods can be grouped by their learning framework, which is either a traditional statistical learner such as SVM (e.g., LIWC) or a neural network (e.g., Text-CNN and SAFE). Baselines can also be grouped as single-modal methods (e.g., LIWC, RST, and Text-CNN) or multimodal methods (e.g., SAFE).

\paragraph{LIWC~\cite{pennebaker2015development}}\footnote{\url{https://liwc.wpengine.com/}}
LIWC (Linguistic Inquiry and Word Count) is a widely-accepted psycholinguistic lexicon. Given a news story, LIWC can count the words in the text falling into one or more of 93 linguistic, psychological, and topical categories, based on which 93 features are extracted and often classified within a traditional statistical learning framework~\cite{zhou2019content}.

\paragraph{RST} RST (Rhetorical Structure Theory) organizes a piece of content as a tree that captures the rhetorical relation among its phrases and sentences. We use a pretrained RST parser~\cite{ji2014representation}\footnote{\url{https://github.com/jiyfeng/DPLP}} to obtain the tree for each news article and count each rhetorical relation (in total, 45) within a tree, based on which 45 features are extracted and classified in a traditional statistical learning framework.

\paragraph{Text-CNN~\cite{kim2014convolutional}} Text-CNN relies on a Convolutional Neural Networks for text classification, which contains a convolutional layer and max pooling. 

\paragraph{SAFE~\cite{zhou2020multimodal}}\footnote{\url{https://github.com/Jindi0/SAFE}} SAFE is a neural-network-based method that utilizes news multimodal information for fake news detection, where news representation is learned jointly by news textual and visual information along with their relationship.
SAFE facilitates recognizing the news falseness in
its text, images, and/or the ``irrelevance'' between the text and images.

\subsection{Implementation Details}
\label{subsec:implementaion_details}

The overall dataset is randomly divided into training and testing datasets with a proportion of 0.8:0.2. As the dataset has an unbalanced distribution between reliable and unreliable news articles ($\approx$2:1), we evaluate the prediction results in terms of precision, recall, and the $F_1$ score. For methods relying on traditional statistical learners, multiple well-established classifiers are adopted in our experiments: Logistic Regression (LR), Na\"ive Bayes (NB), $k$-Nearest Neighbor ($k$-NN), Random Forest (RF), Decision Tree (DT), and Support Vector Machines (SVM). We merely present the one performing best due to the space limitation. Codes are all available on \url{http://coronavirus-fakenews.com}.

\begin{table}[t]
\centering
\caption{Baselines Performance in Predicting COVID-19 News Credibility Using $\mathsf{ReCOVery}$ Data}
\label{tab:baselines}
\begin{tabular}{lcccccc}
\toprule[1pt]
\multirow{2}{*}{\textbf{Method}} & \multicolumn{3}{c}{\textbf{Reliable news}} & \multicolumn{3}{c}{\textbf{Unreliable news}} \\ \cline{2-7}
 & \textbf{Pre.} & \textbf{Rec.} & \textbf{$F_1$} & \textbf{Pre.} & \textbf{Rec.} & \textbf{$F_1$} \\ \midrule[0.5pt]
\textbf{LIWC+DT} & 0.779 & 0.771 & 0.775 & 0.540 & 0.552 & 0.545 \\
\textbf{RST+DT} & 0.721 & 0.705 & 0.712 & 0.421 & 0.441 & 0.430 \\
\textbf{Text-CNN} & 0.746 & 0.782 & 0.764 & 0.522 & 0.472 & 0.496 \\
\textbf{SAFE} & 0.836 & 0.829 & 0.833 & 0.667 & 0.677 & 0.672  \\
\bottomrule[1pt]
\end{tabular}
\end{table}

\subsection{Experimental Results}
\label{subsec:performance}

Prediction results are provided in Table \ref{tab:baselines}. We observe that four baselines achieve an $F_1$-score (precision, recall) score of 71\% (72\%, 71\%) to 83\% (84\%, 83\%) in identifying reliable news and between 43\% (42\%, 44\%) to 67\% (67\%, 68\%) for unreliable news. Additionally, multimodal features are generally more representative than single-modal features in predicting news credibility. We point out that the four baselines are content-based methods; developing more advanced methods by mining social media~\cite{zafarani2014social} are encouraged.

\section{Conclusion}
\label{sec:conclusion}

To fight the coronavirus infodemic, we construct a multimodal repository for COVID-19 news credibility research, which provides textual, visual, temporal, and network information regarding news content and how news spreads on social media. The repository balances data scalability and label accuracy. To facilitate future studies, benchmarks are developed and their performances are presented on predicting news credibility using the data available in the repository. 
We point out that the data could be further enhanced (1) by including COVID-19 news articles in various languages such as Chinese, Russian, Spanish, and Italian, as well as the information on how these news articles spread on the popular local social media for those languages, e.g., Sina Weibo (China). 
Furthermore, (2) extending the dataset by introducing the ground truth of, for example, hate speech, clickbaits, and social bots~\cite{ferrara2019history} would help study the bias and discrimination bred by the virus, as well as the correlation among all information and accounts with low credibility. 
Both (1) and (2) will be our future work.

\section*{Acknowledgments}
Emilio Ferrara is supported by the Defense Advanced Research Projects Agency (DARPA, grant number W911NF-17-C-0094).

\balance
\bibliographystyle{ACM-Reference-Format}
\bibliography{references}

\end{document}